\documentclass[%
 amsmath,amssymb, twocolumn,
 aps, physrev, nofootinbib,
]{revtex4-2}

\usepackage[shadow]{todonotes}	
\usepackage{color, soul}	
\usepackage{wrapfig}	
\usepackage{verbatim}	
\usepackage[normalem]{ulem} 
\usepackage{tabularx} 
\usepackage{graphicx}
\usepackage{cancel}
\usepackage{duckuments} 
\usepackage{txfonts}

\newcommand{\fermi}{\emph{Fermi}}

\newcommand{\orcid}[1]{\href{#1}{\includegraphics[scale=0.035]{figures/orcid-ID.png}}}

\definecolor{boxback}{HTML}{DAE5F0}
\definecolor{boxframe}{HTML}{0F365B}
{
\end{tcolorbox}   
}

\newcommand{\eg}[1]{(e.g. \citealt{#1})}

\def\code#1{\texttt{#1}}

\def\apjl{Astrophys.\ J.\ Lett.}
\def\mnras{Mon.\ Not.\ R.\ Astron.\ Soc.}
\def\araa{Annu.\ Rev.\ Astron.\ Astrophys.}
\def\aap{Astron.\ Astrophys.}
\def\apjs{Astrophys.\ J.\ Suppl.}
\def\pasj{Publ.\ Astron.\ Soc.\ Jpn.}

\usepackage{graphicx}
\usepackage{adjustbox}
\usepackage{dcolumn}
\usepackage{bm}
\usepackage{hyperref}
\usepackage[english]{babel}

\renewcommand{\selectlanguage}[1]{}
\begin{document}

\preprint{APS/123-QED}

\title{\textbf{Black hole spectral states revealed in GRMHD simulations with texture memory accelerated cooling}}

\author{Pedro Naethe Motta}
 \affiliation{Instituto de Astronomia, Geof\'{\i}sica e Ci\^encias Atmosf\'ericas, Universidade de S\~ao Paulo, S\~ao Paulo, SP 05508-090, Brazil.}
 \email{pedronaethemotta@usp.br}

\author{Jonatan Jacquemin-Ide}
\affiliation{JILA, University of Colorado and National Institute of Standards and Technology, 440 UCB, Boulder, CO 80309-0440, USA}

\author{Rodrigo Nemmen}
\affiliation{Instituto de Astronomia, Geof\'{\i}sica e Ci\^encias Atmosf\'ericas, Universidade de S\~ao Paulo, S\~ao Paulo, SP 05508-090, Brazil}
\affiliation{Kavli Institute for Particle Astrophysics and Cosmology, Stanford University, Stanford, CA 94305, USA}

\author{M. T. P. Liska}
\affiliation{Institute for Theory and Computation, Harvard University, 60 Garden Street, Cambridge, MA 02138, USA}
\affiliation{Center for Relativistic Astrophysics, Georgia Institute of Technology, Howey Physics Bldg, 837 State Street NW, Atlanta, GA 30332, USA}

\author{Alexander Tchekhovskoy}
\affiliation{Center for Interdisciplinary Exploration \& Research in Astrophysics (CIERA), Physics \& Astronomy, Northwestern University, Evanston, IL 60202, USA}
\affiliation{NSF-Simons AI Institute for the Sky (SkAI), 172 E. Chestnut St., Chicago, IL 60611, USA}

\date{\today}

\begin{abstract}
X-ray binaries (XRBs) display spectral state transitions that are accompanied by substantial changes in the hardness, luminosity, and structure of the accretion flow. We developed a GPU-accelerated cooling toolkit for general relativistic magnetohydrodynamic (GRMHD) simulations of accreting black holes that uses texture memory for fast retrieval of pre-computed values. The toolkit incorporates bremsstrahlung, synchrotron, inverse Compton radiation and Coulomb collision processes. We implemented our toolkit into a GRMHD code and used it to simulate a magnetically arrested disk in the context of the XRB low/hard state around a Kerr black hole. We explored the mass accretion rate in the $\sim (10^{-6}-0.3) \dot{M}_{\rm Edd}$ range, where $\dot{M}_{\rm Edd}$ is the Eddington accretion rate. Our simulations reveal that for low accretion rates ($\dot{M} \lesssim 0.01 \dot{M}_{\rm Edd}$), the flow settles into a geometrically thick, low-density, two-temperature hot accretion flow. At higher accretion rates, the flow turns into a cold single-temperature thin disk at $r_{\rm in} \gtrsim 50 r_g$. Inside, the disk breaks up into single-temperature thin filaments embedded into a two-temperature hot thick flow. Our GPU texture memory accelerated cooling prescription is $3-5$ times faster than the standard radiation M1 closure methods, and $\sim5$ times faster than storing the lookup table in global memory.
\end{abstract}

\maketitle

\section{Introduction} \label{sec:intro}

Observations of black hole X-ray binary systems (XRBs) reveal changes in observed spectral and timing properties linked to spectral state transitions, which occur cyclically over periods ranging from months to years \citep{Esin_1997, Remillard_2006, Fender_2012}. XRBs oscillate between three states. In the hard or low state, their X-ray spectra are described by a power-law with photon index of $\approx 1.6 - 2.1$ with a cutoff around tens of keV. Radio emissions associated with steady jets are also commonly observed in the hard state \eg{Corbel_2002, Fender_2004}. As it brightens, it transitions into the high or soft state, where the spectrum is dominated by a thermal component. Additionally, this state is associated with the quenching of the jets due to the disappearance of steady radio/IR emission \citep{Coriat_2009, Coriat_2011}. A smooth transition between these two states characterizes an intermediate state (very high state) where thermal and power-law components of the spectrum are comparable.

The widely accepted scenario, proposed by \citet{Esin_1996}, suggests that changes in the geometry of the accretion flow are responsible for the spectral states. The model proposes that a hot, geometrically thick corona, described as a radiatively inefficient accretion flow (RIAF) \citep{Yuan_2014}, is present in the inner region of the disk, transitioning to an outer, colder geometrically thin disk \citep{Shakura_1973, Novikov1973} at a truncation radius $r = r_{\rm in}$. In the hard state, the disk is truncated at large distances ($r_{\rm in} \rightarrow \infty$). As the mass accretion rate ($\dot{M}$) increases, $r_{\rm in}$ gradually decreases until it reaches the innermost stable circular radius and the system reaches the soft state. 

A key question arising from the proposed scenario is: What mechanism is responsible for the formation and stabilization of the corona? The presence of a geometrically thin disk is well accepted \citep{Done_2007}, but the physical properties of the inner RIAF remain a topic of debate. For instance, there is an ongoing discussion as to whether the state transition mechanism is driven by a decrease in the inner edge of the thin disk \citep{Fender_2004, Ingram_2011, Plant_2014} or a reduction in the size of the corona \citep{Garcia_2015, kara_2019, Wang_2021}.

\citet{ferreira_unified_2006} proposed that a strong large-scale vertical magnetic field is crucial for forming and stabilizing a corona-like inner disk. This large-scale vertical magnetic field is also an essential part of jet launching, as all jet launching mechanisms require it \citep{Blandford1977,Blandford1982}. Thus, the transport of this magnetic field impacts the spectral state evolution of the accretion disk as well as its radio emission. Follow-up studies using a 1D model based on this framework have successfully reproduced the spectral evolution of the X-ray binary GX 339-4 \citep{marcel_unified_2018,marcel_unified_2018-1,marcel_unified_2019}. However, the underlying dynamics, including magnetic field transport and its role in thermodynamics, remain poorly understood.
Fully three-dimensional simulations of magnetized accretion disks are essential to comprehend the establishment of an inner disk corona, its relationship to jet launching and the quintessential role of the large-scale vertical field.

To properly understand the dynamics and thermodynamics of the plasma in strong gravitational field regime, we rely on general relativistic magnetohydrodynamical simulations (GRMHD). However, simulations of the formation and dynamics of truncated disks are challenging because at the relevant luminosity regime, radiation should play a crucial role in the evolution of the system and its impact needs to be incorporated. This is not trivial \eg{Davis2020}. Also, the thinner the disk, the higher the resolution needed to resolve the magnetic turbulence that drives the angular momentum transport. Finally, the viscous time scale of thin disks is $\sim 10^7 r_g/c$ \citep{Nemmen2024}, adding another layer of computational cost to adequately simulate these systems.

In recent years, the Magnetically Arrested Disk (MAD) has emerged as a promising configuration for explaining highly magnetized jetted systems \citep{Narayan2003,Igumenshchev2003,Tchekhovskoy_2011}. A MAD represents the end state of an accretion disk with a sufficiently large reservoir of magnetic flux, this flux can be transported inward from larger scales \citep{tchekhovskoy_magnetic_2015,jacquemin-ide_magnetic_2021} or generated within the accretion disk \citep{liska_large-scale_2020,jacquemin-ide_magnetorotational_2024}. As the magnetic field is transported inward by the accretion flow, it saturates the black hole with enough flux to obstruct gas infall and enable efficient jet launching.

Many radiation MHD simulations have been conducted in the context of X-ray binaries (XRBs). \citet{Takahashi_2016} performed general relativistic radiation magnetohydrodynamics (GRRMHD) simulations using the M1 closure radiation scheme at an accretion rate of $\dot{m} = 0.3$. \citet{Liska_2022} conducted GRRMHD simulations with both toroidal and poloidal magnetic field initial conditions, finding that only the MAD poloidal configuration developed a truncated disk. They suggested that magnetic pressure serves as the truncation mechanism, preventing disk collapse in the inner regions. Similarly, \citet{Dexter_2021} studied a MAD flow using GRRMHD simulations. While they observed disk collapse in one simulation, they did not report a truncated disk. See also the recent work by \citet{hankla_inner_2025}, who report a truncated accretion disk in a weakly magnetized regime, with a truncation radius that depends on the Eddington ratio.
Lastly, \citet{Nemmen2024} performed 2D hydrodynamical simulations for accretion rates ranging from $0.02$ to $0.35$, using a cooling prescription similar to the one used in this work. Their results showed a truncated cold disk embedded in a hot corona, with the inner radius of the cold disk following $r_{\rm{in}} = 42r_g \left(\dot{m}/0.1\right)^{-0.60}$.

In this work, we introduce a computationally-efficient approach for including more physics in the cooling processes relevant to black hole accretion flows, more specifically, we use a radiative prescription source term \citep{Narayan_1995, Esin_1996} and store it as a lookup table in the GPU texture memory. We then present the results of five simulations covering a range of accretion rates, $\dot{M} = (3 \times 10^{-6} - 0.26)\ \dot{M}_{\rm {Edd}}$,  using this GPU-accelerated cooling prescription implemented in the \code{H-AMR} code \citep{liska_2022_hamrcode}---a GPU-accelerated, finite volume, shock capturing, Godunov-based Harten-Lax-van Leer-Einfeldt (HLLE) scheme. We focus on simulating MADs as they are more likely to develop a truncated disk structure due to the presence of a strong and large-scale vertical magnetic field \citep{ferreira_unified_2006,Liska2022}.

This paper is organized as follows: Section~\ref{sec:theoretical_equations} introduces the cooling physics used in our simulations. Section \ref{sec:numerical_methods} details the advantages of utilizing texture memory and the integration of the cooling equations into the \code{H-AMR} framework. In Section~\ref{sec:results}, we present the findings of the five simulations performed, followed by a discussion of these results in Section~\ref{sec:discussion}. Finally, Section~\ref{sec:conclusion} summarizes our key results and outlines potential directions for future research.

\section{Cooling physics}   \label{sec:theoretical_equations}

In this Section, we describe the emission processes included in our cooling table. We compute the total cooling rate that accounts for the effect of bremsstrahlung, synchrotron and a local Compton effect for the synchrotron radiation. We then use a bridge function to transition the cooling to the black-body emission when the medium is optically thick \citep{Esin_1996,Narayan_1995}. 

The intermediate state in XRBs includes plasma that can be both optically thin and thick. To model this, we utilized the two-temperature plasma module in H-AMR~\citep{Ressler_2015, Liska_2022}, which enables separate evolution of ion and electron entropy tracers by distributing the dissipated energy between the two fluid components. We also include Coulomb collisions to account for energy exchange between electrons and ions.

\subsection{Bremsstrahlung emission}
The bremsstrahlung prescription is described as $Q^-_{\rm brem} = Q^-_{\rm ee} + Q^-_{\rm ei}$, where the electron-electron $(Q^-_{\rm ee})$ and electron-ion $(Q^-_{\rm ei})$ modeled interactions are: 
\begin{align}
    Q_{\rm ee}^- &= 2.56\times10^{-22}n_{\rm{e}}^{2} \;{\rm [ergs\ cm^{-3}\ s^{-1}]}\notag \\
    &\quad \times 
    \begin{cases}
        \theta_{\rm e}^{3/2} \left(1+ 1.1 \theta_{\rm e} + \theta_{\rm e}^2 - 1.25 \theta_{\rm e}^{5/2} \right), 
        & \text{if } \theta_{\rm e} < 1 \\
        1.34 \theta_{\rm e} \left[\ln(1.123\theta_{\rm e}) + 1.28\right], 
        & \text{if } \theta_{\rm e} \ge 1
    \end{cases}\\
    Q^{-}_{\rm ei} &= 1.48 \times 10^{-22} n_{\rm e}^2 \;{\rm [ergs\ cm^{-3}\ s^{-1}]} \notag \\
    &\quad \times
    \begin{cases}
        4\left(\displaystyle\frac{2\theta_{\rm e}}{\pi^3}\right)^{1/2}[1 + 1.781\theta_{\rm e}^{1.34}], & \text{if $\theta_{\rm e} < 1$} \\
        \displaystyle\frac{9\theta_{\rm e}}{2\pi}[\ln(1.123\theta_{\rm e} + 0.48) + 1.5], & \text{if $\theta_{\rm e}\ge1$}
    \end{cases}
\end{align}
and $\theta_{\rm e} = k_{\rm B} T_{\rm e}/m_{\rm e} c^2$ is the dimensionless electron temperature. Here, $c$ is the speed of light, $n_{\rm e}$ is the electron density $k_{\rm B}$ is the Boltzmann constant, $T_{\rm e}$ is the electron temperature and $m_{\rm e}$ is the electron mass.

\subsection{Synchrotron emission}
The synchrotron emission will be important within the corona, where the disk is very hot and electrons are accelerated to relativistic speeds. According to \citet{Pacholczyk_1970}, for an isotropic Maxwellian velocity distribution of electrons the emissivity distribution is calculated to be
\begin{equation}
        \epsilon_s d\nu = 4.43 \times 10^{-30} \frac{4 \pi \nu n_{\rm e}}{K_2(1/\theta_{\rm e})} I^\prime\left(x_M\right) d\nu \ \rm{[erg \ cm^{-3} \ s^{-1}]}
        \label{optically_thin_synch}
\end{equation}
where $x_M = (4 \pi m_{\rm e} c\nu)/(3 e^- B \theta_{\rm e}^2)$ with $B = \lvert \Vec{B} \rvert$ being the magnitude of the magnetic field. Here, $\nu$ is the frequency of the emitted photon, $K_2$ is the modified Bessel function of the second kind, $e^-$ is the electron charge and $I^\prime (x_M)$ is a function given by \citep{Mahadevan_1996}
\begin{equation}
    I^\prime(x_M) = \frac{4.0505}{x_M^{1/6}} \left(1 + \frac{0.4}{x_M^{1/4}} + \frac{0.5316}{x_M^{1/2}}\right)\exp\left(-1.8899x_M^{1/3}\right).
    \label{fitting_I(xm)}
\end{equation}
The frequency integration of $\epsilon_s$ yields the synchrotron emission.  However, below a critical frequency $\nu_c$, the emission becomes self-absorbed and it is expected to reproduce a black body spectrum. We can estimate the critical frequency considering the synchrotron emission from a thin annulus of thermal scale height $2H_{\rm th}$, radius $R$ and thickness $\Delta R$ to be equal to the blackbody emission in the Rayleigh-Jeans limit from the upper and lower surface of the annulus. This yields
\begin{gather}
   {  2H_{\rm th}(2\pi R \Delta R) }  \times
   \epsilon_s d\nu   =
   {   2   } \times 
   {(2 \pi R \Delta R)} \times
   {\frac{2 \pi \nu_c^2 k_{\rm B} T_{\rm e}}{c^2}} d\nu.
   \label{bb_sync_aprox}
\end{gather}
Given $H_{\rm th}$, $R$ and $T_{\rm e}$ and using equations \eqref{optically_thin_synch} and \eqref{fitting_I(xm)} in  \eqref{bb_sync_aprox}, we can obtain $\nu_c$ by numerical integration. Then, we consider any emission below $\nu_c$ to be self-absorbed, so that the volume emissivity can be approximated by the blackbody emission from the surface of the disk, divided by the disk volume. Above $\nu_c$, the emission is optically thin and can be approximated by the frequency integration of equation \eqref{optically_thin_synch}. To get the total cooling per unit volume, we integrate over the frequency
\begin{equation}
    Q_{\rm syn}^{-} = \frac{2\pi R^2}{2H_{\rm th}\pi R^2} \int_0^{\nu_c} 2\pi \frac{\nu^2}{c^2} k_{\rm B} T_{\rm e} d\nu + \int^{\infty}_{\nu_c} \epsilon_s (\nu) d\nu.
    \label{synchrotron}
\end{equation}
We encounter a problem when calculating the synchrotron emission rate for low temperatures $(T_{\rm e} \leq 10^8 K)$ because the modified Bessel function of the second kind approaches zero $K_2(1/\theta_{\rm e})$. To overcome this issue, we follow \citet{Fragile_2009} and substitute it for $2 \theta_{\rm e}^2$ in low-temperature regimes. 

\subsection{Inverse Compton enhancement}
\label{Inverse Compton enhancement}
Relativistic electrons in the corona usually transfer energy to photons emitted by the disk by means of the inverse Compton effect, which leads to an enhancement of the synchrotron emission by a factor $\eta_{\rm comp}$. Our approach is to consider a local Compton effect described by an energy enhancement factor $\eta_{\rm comp} =$ $ \langle \Delta E \rangle /E_{\rm init}$ \citep{Dermer_1991}, where $\langle \Delta E \rangle$ is the averaged change in energy of a photon from the moment it enters the medium until it escapes and $E_{\rm init}$ is its initial energy. It is possible to describe the Compton enhancement factor by
\begin{gather}
    \eta_{\rm comp} = 1 + \frac{P(A-1)}{1 - PA} \left[1 - \left(\frac{E_{\rm init}}{3\theta_{\rm e}}\right)^{1 - \ln P/\ln A}\right];
    \\
    P = 1 - \exp({-\tau_{\rm sca}});
    \\
    A = 1 + 4 \theta_{\rm e} + 16\theta_{\rm e}^2,
\end{gather}
where $P$ is the probability of scattering the photon, $A$ the average factor by which the photon energy increases, $\tau_{\rm sca} = 2n_{\rm e}\sigma_{\rm T} H_{\rm th}$ is the Thomson optical depth, and $\sigma_{\rm T}$ is the Thomson cross section. Note that the formula above is only valid for soft photons, e.g. $E_{\rm{init}} < 3\theta_{\rm e}$, harder photons will heat the electrons, but we do not encounter this effect in the ranges considered here.

In semi-analytic accretion models, the synchrotron spectrum is strongly peaked at $\nu_c$ \citep{Narayan_1995}. To simplify the cooling calculation, we approximate the Compton effect by setting $\eta_{\rm comp}(\nu) \approx \eta_{\rm comp}(\nu_c)$. The total optically thin cooling of the disk is finally given by
\begin{equation}
        Q^{-}_{\textit{\rm thin}} (H_{\rm th}, B, n_{\rm e}, T_{\rm e}) = Q_{brem}^{-} + \eta_{\rm comp}(\nu_c)Q_{syn}^-.
    \label{thin_cooling}
\end{equation}

\subsection{Optically thick cooling}

We aim to understand and implement the effect of cooling during the intermediate state where both optically thin and thick regimes coexist. We extend the prescriptions defined above, valid for the optically thin limit, using the following approximation for the total cooling \citep{Hubeny_1990},
\begin{equation}
    Q^{-}_{\textit{\rm total}} (H_{\rm th}, B, n_{\rm e}, T_{\rm e}) = \frac{4\sigma_{\rm T} T_{\rm e}^4}{H_{\rm th}}\frac{1}{3\tau/2 + \sqrt{3} + 1/\tau_{\rm abs}} 
    \label{total_cooling}
\end{equation}
where $\tau = \tau_{\rm abs} + \tau_{\rm sca}$ is the total optical depth, $\tau_{\rm abs}$ is the absorption optical depth and $\tau_{\rm sca}$ is the Thomson optical depth defined in Section \ref{Inverse Compton enhancement}. We approximate $\tau_{\rm abs}$ as \citep{Narayan_1995} 
\begin{equation}
    \tau_\text{abs} = \frac{H_{\rm th}}{4 \sigma_{\rm T} T_{\rm e}^4} Q^{-}_{\textit{\rm thin}}.
\end{equation}
Note that eq. \eqref{total_cooling} is valid for both regimes: when $\tau \gg 1$ (optically thick), it yields the blackbody radiation limit, and when $\tau \ll 1$ (optically thin), eq. \eqref{total_cooling} reduces to eq. \eqref{thin_cooling}.

\subsection{Parameters}
\label{subsec:Parameters}
Our approximation for cooling, $Q \equiv Q(H_{\rm th}, B, n_{\rm e}, T_{\rm e})$, depends on four parameters: Disk thermal scale height, magnetic field strength, electron density and electron temperature, respectively. We implemented the cooling function in the \code{H-AMR} code using a lookup table. The parameters are sampled in cgs units with $100$ points, equally spaced in log space: $H_{\rm th} \in [10^3, 10^{12}] \ [\rm{cm}], B \in [1, 10^{10}] \ [\rm{G}], n_{\rm e} \in [10^2, 10^{25}] \ [\rm{cm^{-3}}]$ and $T_{\rm e} \in [10^2, 10^{15}]\  [\rm{K}]$. This results in a table with $100^4$ values. We then used the specific in-cell values of each parameter to calculate the cooling rate for each cell.

\subsection{Disk scale height}
Most quantities, such as \( n_{\rm e} \), \( T_{\rm e} \), and \( B \), are easily computed locally at each cell, but the disk scale height, \( H_{\rm th} \), is not. The scale height depends on the global disk structure and cannot be calculated using a single cell, creating a challenge in determining the local cooling within a cell. 
We deal with this problem by leveraging the relationship of the disk scale height with the thermal equilibrium of the disk. For most accretion disks the scale height is,
\begin{equation}
    H_{\rm th} = {c_s/\Omega_{\rm K}},
    \label{eq:H_thermal}
\end{equation}
where $c_s$ is the relativistic sound speed and $\Omega_{\rm K}$ is the Keplerian frequency. To calculate $c_s$, we follow
\begin{equation}
    c_s = \sqrt{\frac{\gamma P}{\rho h}},
\end{equation}
where $P$ is the gas pressure, $\rho h$ is the relativistic enthalpy density given by $\rho h = \rho c^2 + \rho \epsilon + P$, $\rho \epsilon$ is the gas internal energy density and $\gamma$ is the adiabatic index. Additionally, the Keplerian frequency can be calculated as $\Omega_{\rm K} =1/(r^{3/2} + a)$ in natural units. 

\subsection{Coulomb Collisions}
\label{subsec:Coulomb_Collisions}
We have based our implementation on the equations described by \citet{sadowski_2016, Liska_2022}, which has shown itself to be a reliable approach for modeling Coulomb collisions in similar systems,
\begin{equation}
    \begin{split}
        Q_{\rm cc} = &\frac{3}{2}\frac{m_{\rm e}}{m_{\rm i}} \frac{\rho}{m_{\rm p}}n_{\rm e} \log(\Lambda) \frac{c k_{\rm B} \sigma_{\rm T} (T_{\rm i} - T_{\rm e})}{K_2(1/\theta_{\rm i}) K_2(1/\theta_{\rm e})}
         [{\rm erg \; cm^{-3} s^{-1}}]
        \\
        &\times \left[ \frac{2(\theta_{\rm e} + \theta_{\rm i})^2 + 1}{\theta_{\rm e} + \theta_{\rm i}} K_1 (1/\theta_{\rm m}) + 2K_0 (1/\theta_{\rm m})\right],
    \end{split}
    \label{eq:coulomb_collisions}
\end{equation}
where $\Lambda \approx 20$, $K_0, K_1$ and $K_2$ are different kinds of modified Bessel functions, $\theta_{\rm m} = (1/\theta_{\rm e} + 1/\theta_{\rm i})^{-1}$, and $\theta_{\rm i} = k_{\rm B}T_{\rm i}/m_{\rm e}c^2$. We calculate the Coulomb collision heating rate density as a function of three variables, $Q_{\rm cc}(n_{\rm e}, T_{\rm i}, T_{\rm e})$. We pre-compute these values in a lookup table that uniformly samples in the log space with 100 points the following variables over the indicated ranges: $n_{\rm e} = [10^2, 10^{25}], T_{\rm i} \in [10^2, 10^{15}], T_{\rm e} \in [10^2, 10^{15}]$, all measured in cgs units. This results in a table of $100^3$ values. 

\section{Numerical Methods}
\label{sec:numerical_methods}

\subsection{GPU texture memory}    

To maximize the computational efficiency of the cooling model, we implement the look-up table using GPU texture memory.
Texture memory is a read-only, cached on-chip, digital storage that makes texture data easily available to the GPU. \citet{schneider2017} found a considerable increase in efficiency using this method to implement radiative cooling for simulating the galactic winds in the \code{CHOLLA} code~\citep{schneider2015}. Because texture memory is cached on-chip, usually it provides higher effective bandwidth in comparison to requesting memory access to the off-chip DRAM. Aside from that, texture memory also has a great sense of spatial locality, which means that the data for adjacent texels (texture elements) are stored close together in memory. This can help to improve the performance of graphics rendering, as the GPU can access the data more quickly and efficiently, reducing the time it takes to retrieve it from memory. Texture memory features a linear interpolation method for each dimension, which can help us to deal with in-between table values. In Appendix~\ref{Apendix_sec: efficiency}, we compare the relative speedup of using the lookup table in texture memory versus global memory, as well as computing the cooling rate on the fly, both within and outside of the \code{H-AMR} framework.

Texture memory was originally developed for graphical purposes, which makes it possible to store data up to 3D arrays. Therefore, we use two 3D texture objects to store values from the cooling and coulomb tables. Each dimension of the texture memory is set to be one of the parameters, but since the cooling values depend on 4 parameters, we had to flatten two dimensions of the original array into one. We chose to flatten $n_{\rm e}$ and $T_{\rm e}$.

The flattening process is straightforward. Following a single texture memory dimension, we save the cooling values in each texel as we increase the temperature. After running through all $T_{\rm e}$ values, we move on to the next value of $n_{\rm e}$ and calculate the cooling again for every $T_{\rm e}$ value. We follow this approach until we cover the entire $n_{\rm e}$ space. 

Although this process is simple, the interpolation done by the texture memory does not work in this dimension. Hence, we manually interpolate this dimension while the other two are done by the texture memory. 

\subsection{Implementing realistic cooling into \code{H-AMR}}  \label{sec:implementation_in_hamr}

We implement our cooling function into the \code{H-AMR} code, by evolving the following energy-momentum equation \citep{Mckinney_2014},
\begin{equation}
        \nabla_\mu T^\mu_{\ \alpha} = G_\alpha,
    \label{energy-momentum_eq}
\end{equation}
where 
\begin{equation}
    T^\mu_{\ \alpha} = (\rho + u_{\rm g} + p_{\rm g} + b^2) u^\mu u_\alpha + (p_{\rm g} + \frac{b^2}{2}) \delta^\mu_{\ \alpha} - b^\mu b_\alpha
\end{equation}

is the mixed GRMHD stress-energy tensor, $b^\mu$ is the magnetic field 4-vector, $p$ is the gas pressure, $u_g$ is the gas internal energy, $u^\mu$ is the 4-velocity, $\delta^\mu_\nu$ is the Kronecker delta function and $G_\nu$ is the external 4-force density. Here and below, we absorb the factor of $(4\pi)^{-1/2}$ into the definition of the magnetic field. We do not consider explicit radiation pressure and only account for the local radiative energy loss in each cell of the grid. However, our thermal pressure could model this radiation pressure at the cost of inaccurate temperatures in the optically thick regions. We discuss this approximation and correct the temperature structure in section \ref{sec:thermalstructure}.

We consider the 4-force density as:
\begin{equation}
    G_\mu = - \lambda u_\mu.
    \label{4-force_calc}
\end{equation}
In the fluid frame, this reduces to $G_t = \lambda$, where $\lambda = Q^-_{\text{total}}$ is the energy loss rate density due to radiative cooling, as given by eq.~\eqref{total_cooling}.

The cooling function utilizes the values of $H_{\rm th}$, $B$, $n_{\rm e}$, and $T_{\rm e}$ obtained from the simulation to find the appropriate interpolated cooling rate from a precomputed table stored in the GPU texture memory as described in Section~\ref{subsec:Parameters}. It then applies this cooling rate to the calculation of the 4-force density in equation \eqref{4-force_calc} and updates the energy-momentum equation \eqref{energy-momentum_eq}. Similar to our implementation of radiative cooling, for the Coulomb collision energy transfer rate we also use a pre-computed table stored in texture memory as described in Section~\ref{subsec:Coulomb_Collisions}. We determine the Coulomb collision rate by evaluating $n_{\rm e}$, $T_{\rm i}$ and $T_{\rm e}$ from the simulation. The texture function then uses these parameters to obtain the appropriate Coulomb rate from the table. 

\subsubsection{Implicit vs explicit schemes} \label{Handling implicit values}

When cooling is strong, the cooling timescale can become much shorter than the system's dynamical timescale, requiring a very small time step. Additionally, large cooling rates can lead to numerical errors if the associated decrease in energy exceeds the cell internal energy. To manage this, we use two methods: (i)~setting an upper limit for the cooling rate, and (ii)~implementing an implicit solver for the cooling step. For (i), we limit the cooling rate to result in an internal energy update by at most $30\%$.
For (ii), we have adapted the implicit-explicit (IMEX) solver from the \code{H-AMR} code, used in the M1 radiation scheme~\citep{Liska_2022}, to work with our cooling prescription. 
We retain both options, because there are cases where the implicit method can slow down the code. We discuss the differences between the implicit and explicit approaches in Section~\ref{Comparison between implicit and explicit methods}.

\subsubsection{Temperature floor} 

For reasons not yet fully understood, but related to the disk thermal instability \citep{shakura_theory_1976}, radiative simulations of accretion disks at certain accretion rates are prone to runaway cooling, leading to the thermal collapse of the disk and making it impossible to numerically resolve its thickness \citep{jiang_thermal_2013, Liska_2022}. Our cooling approach encounters the same issue. To address this, we implement a temperature floor in our simulations. This floor functions similarly to the cooling prescription of \citet{Noble_2009}, where a target dimensionless scale height $h_{\rm floor} = (H/R)_{\rm floor}$ is used to set the minimum allowed temperature.

In practice, we calculate the gas temperature for each cell and compare it to the temperature floor for a specific $h_{\rm floor}$. The temperature threshold for each cell is given by
\begin{equation}
    T_{\rm floor} = \frac{\pi}{2} \left[\left(\frac{H}{R}\right)_{\rm floor} \Omega_{\rm K} R\right]^2.
\end{equation}
If the temperature falls below the floor value, the cooling rate in that cell is set to zero.
\subsection{Numerical setup}

In all our simulations, we set up initial conditions for MADs because they are more likely to develop a truncated accretion disk structure \citep{Liska2022}.
We first run an uncooled MAD simulation, \code{INIT}, described below, for \( 21,300 \, r_g/c \). This serves as the starting point for all cooled simulations. This approach saves time, as the magnetic field structure settles more quickly in an uncooled MAD. Once \code{INIT} reaches a quasi-steady state out to sufficient radii $(r \sim 40 \ r_g)$, its final snapshot initializes all cooled simulations. The duration of each cooled simulation varies and is detailed in Table \ref{tab:simulation_list}.

The initial condition of the \code{INIT} simulation is initialized with the standard \citet{Fishbone_1976} solution, with \( r_{\rm in} = 20 r_g \) and \( r_{\rm max} = 41 r_g \), where \( r_{\rm max} \) denotes the location of the density maximum.
To avoid magnetic monopoles in the initial condition, we initialize the magnetic fields by setting the initial toroidal vector potential
\begin{equation}
    A_{\varphi}^{\rm MAD} = \frac{\rho}{\rho_{\rm max}} \left(\frac{r \sin{\theta}}{20}\right)^3 e^{-r/400} - 0.2,
    \label{MAD_potential}
\end{equation}
where \( \rho_{\rm max} \) is the density maximum of the \citet{Fishbone_1976} tori. The initial field strength is set by the plasma beta, \( \beta = \frac{P}{(b^2/8\pi)} \)  using \( \beta = 100 \). We set $h_{\rm floor} = 0.1$, the black hole spin $a_* = 0.9375$ and the adiabatic index to \( \gamma = 5/3 \) across all simulations in this paper. Based on \citet{Liska_2024, Chael_2025}, this choice is well-suited for our high-accretion-rate and mid-accretion-rate runs but is not necessarily a good approximation for our low-accretion-rate simulations, where the accretion rate is measured relative to the Eddington accretion rate (see table \ref{tab:simulation_list}).

We use a resolution of $384 \times 300 \times 64$ for the MAD simulations. We show that this resolution is sufficient to resolve the MRI in Appendix \ref{Appendix:quality_factors}. To explore different cooling regimes, we adjust the mass density scale, \( \rho_{\rm scale} \). This scale directly controls the accretion rate relative to the Eddington limit in our simulations, resulting in different cooling regimes within the accretion disk. 

\section{Results}
\label{sec:results}
We performed five simulations with varying mass accretion rates (high, intermediate and low with respect to the Eddington accretion rate) and implicit/explicit cooling schemes. Details of each simulation can be found in Table \ref{tab:simulation_list}. From now on, we will refer to the simulations that use the upper limit for cooling rate as \textit{explicit} and those that evolve the system implicitly as \textit{implicit}, as described in Section \ref{Handling implicit values}.

For each simulation, we define the Eddington accretion rate as,
\begin{equation}
    \dot{M}_{\rm Edd} =  \frac{4 \pi G M m_{\rm p}}{\eta c \sigma_{\rm T}}.
\end{equation}
Here, $m_{\rm p}$ is the mass of a proton and $G$ is the gravitational constant. We consider a black hole mass of $M = 10 M_\odot$ and compute the radiative efficiency defined as
\begin{equation}
    \eta = \frac{\int_0^{2\pi} \int_0^\pi \int_{3 \rm r_g}^{100\rm r_g} \sqrt{-g}\ Q^-_{\rm Total} dr d\theta d\varphi}{\langle\dot{M} c^2\rangle},
    \label{eq:efficiency_value}
\end{equation}
where $\dot{M}$ is the accretion rate defined below. 

The cooling is computed within the code and stored in the dump files. However, due to technical difficulties with the cooling output, we only have the cooling function for the final snapshot of each simulation. We stress that this is merely a dumping issue and the cooling was correctly calculated for the whole duration of the simulation.

We calculate the efficiency considering the last snapshot of each simulation, yielding an efficiency of 0.1494 for the \code{EXP\_HIGH}, 0.1443 for \code{IMP\_HIGH}, 0.0484 for the \code{EXP\_MID}, 0.0066 for the \code{EXP\_LOW} and 0.0044 for the \code{IMP\_LOW}. 

We calculate the accretion rate and the normalized magnetic flux for each simulation as
\begin{gather}
    \dot{M}(r) = - \int^{2 \pi}_{0} \int^\pi_0 \sqrt{- g} \rho u^r d\theta d\varphi,
    \label{accretion_rate}
    \\
    \phi_{\rm BH}  = \frac{1}{2}\frac{1}{\sqrt{\langle\dot{M}\rangle_t|_{5r_g}}}\int^{2 \pi}_{0} \int^\pi_0  \sqrt{- g} \lvert B^r\rvert_{r=r_{\rm BH}} d\theta d\varphi,
    \label{magnetic_flux}
\end{gather}
where we compute the dimensionless magnetic flux at the horizon of the BH and normalize it by the running time-averaged accretion rate over a $18,750\ r_g/c$ time-window \footnote{The choice of $r =5\ r_g$ is to minimize contributions of density floors to the accretion rate.} at $r =5\ r_g$.

\begin{figure}[htpb]
    \centering
    \includegraphics[width= \linewidth]{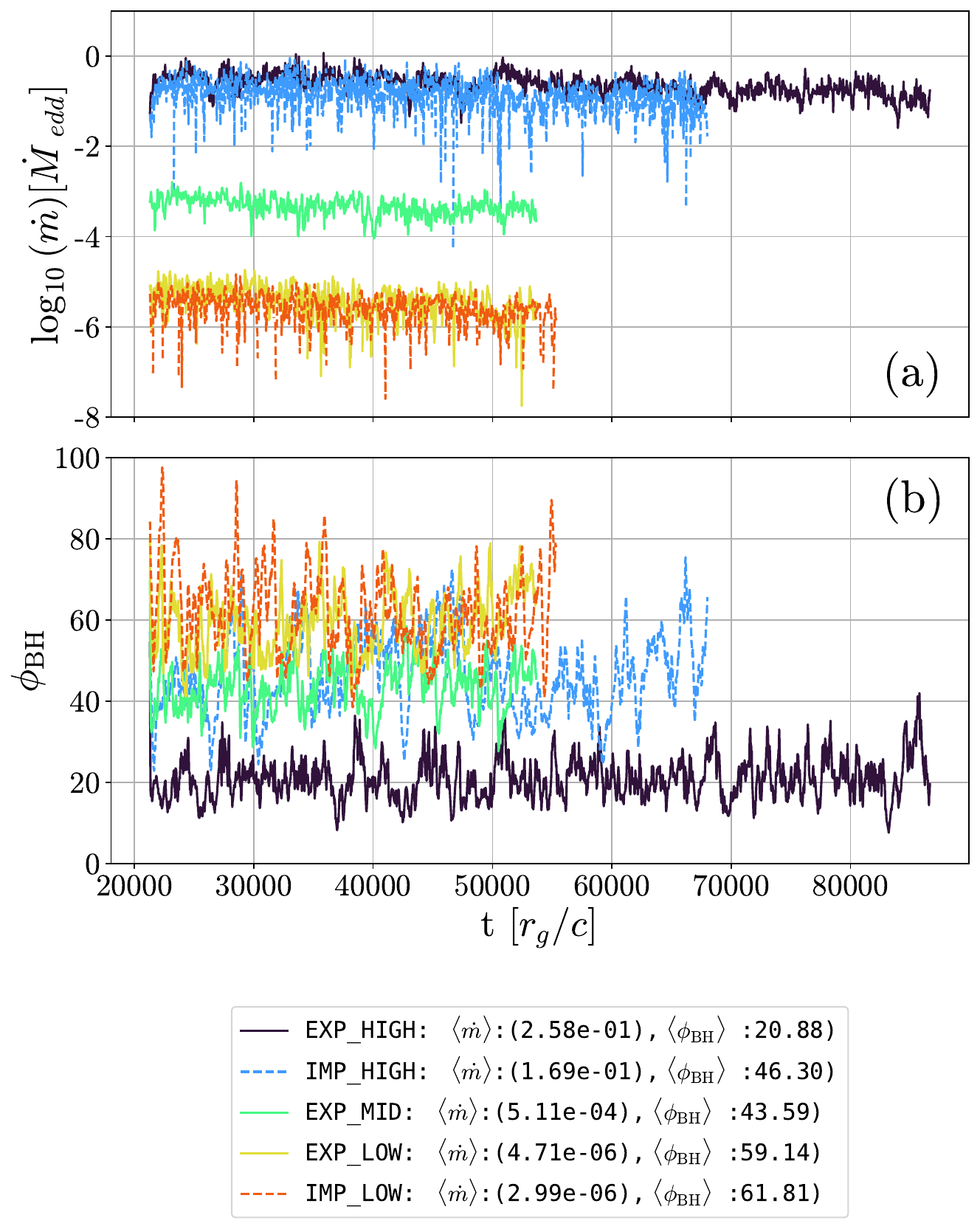}
    \caption{Panel (a): Accretion rate in Eddington units for all simulations as described in equation \eqref{accretion_rate}. Panel (b): Magnetic flux normalized by the square root of the running time average mass accretion rate for all simulations. We average the accretion rate over the entire duration of the cooled simulation. We show the time average values of the accretion rate and magnetic flux for each simulation in the legend.}
    \label{fig:mdot_phimag}
\end{figure}

In Figure \ref{fig:mdot_phimag}(a), we depict the accretion rate normalized to the Eddington accretion rate $\dot{m} = \dot{M}/\dot{M}_{\rm edd}$, where $\dot{M}$ is given by equation \eqref{accretion_rate}. The average Eddington accretion rate values can be found in Table \ref{tab:simulation_list}. We recover a clear separation of the high, mid and low simulations by their Eddington ratio. We notice that all simulations exhibit a quasi-steady state. In the high Eddington case, we observe that the implicit and explicit cooling regimes produce significantly different amplitudes of accretion rate variability, with the implicit simulation showing comparatively larger variability. However, in the low Eddington case, both simulations exhibit similar spikes in accretion variability.

In Figure \ref{fig:mdot_phimag}(b), we plot the dimensionless magnetic flux ($\phi_{\rm BH}$) as defined in equation \eqref{magnetic_flux} at the event horizon as a function of time. We observe a quasi-steady state for all simulations, which signals the presence of a MAD. 

We observe differences in the saturation values and evolution of $\phi_{\rm BH}$ across simulations. Specifically, simulations at high Eddington ratios have, on average, smaller values of \( \phi_{\rm BH} \) compared to those with low Eddington ratios. Additionally, for the high Eddington ratio case, the evolution differs significantly, with a factor of 2 difference in the final \( \phi_{\rm BH} \) value. On the contrary, the magnetic flux, \( \phi_{\rm BH} \), evolves almost identically for \code{EXP\_LOW} and \code{IMP\_LOW}. Showing again that implicit and explicit cooling methods differ the most in the high Eddington accretion rate regime. 
The trend in $\phi_{\rm BH}$ is shown in Section \ref{sec:flow_thin} to be explained by the disk thickness of the different simulations.

\begin{figure}[htpb]
    \centering
    \includegraphics[width= \linewidth]{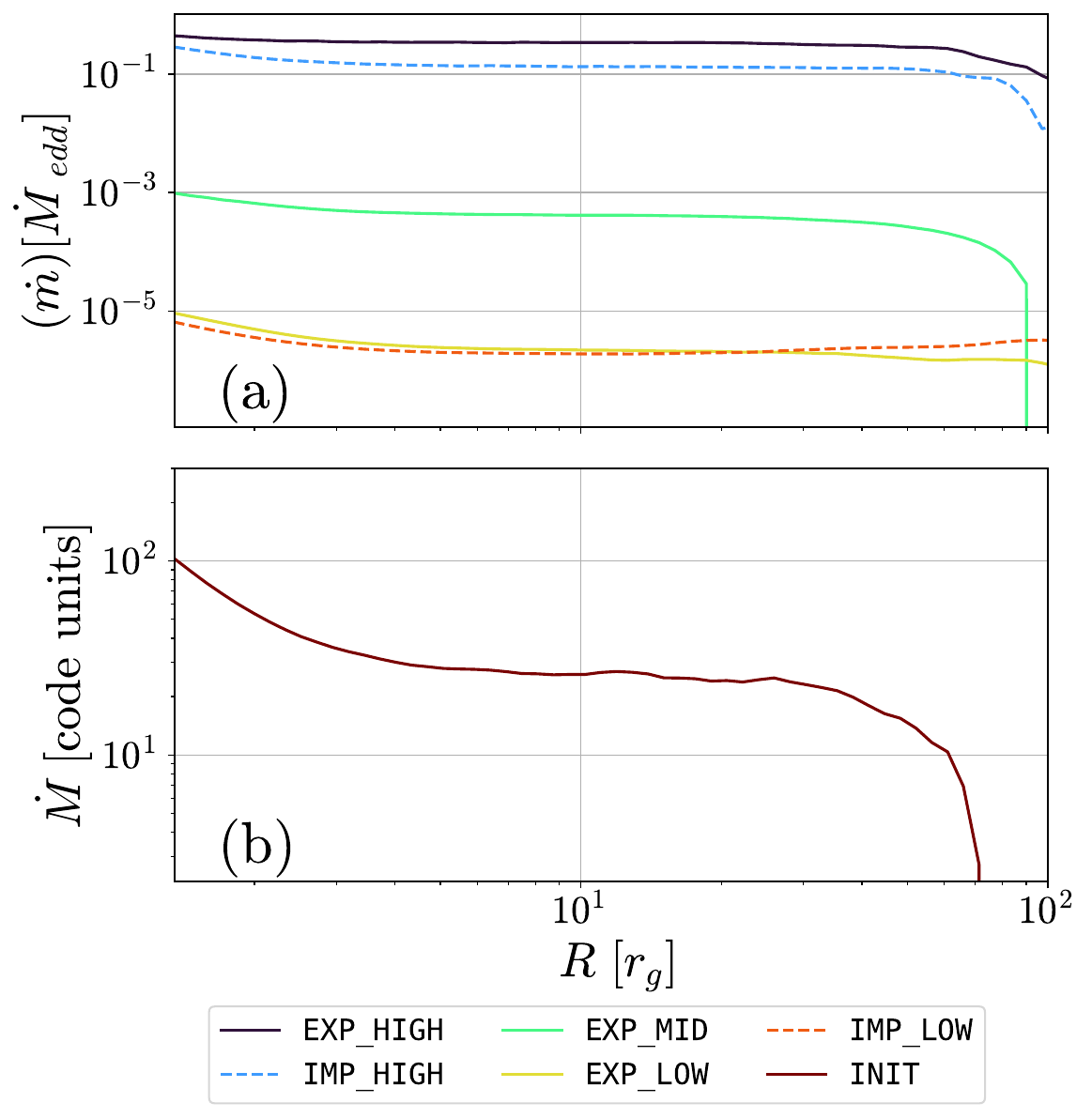}
    \caption{Panel (a): The time-averaged accretion rate, $\dot{M}$, in Eddington units as a function of distance. This is averaged over the interval $48,675$ to $53,675 r_g/c$. Panel (b): The time-averaged accretion rate profile radial profile of the mass accretion rate, in code units, as a function of the distance. This is averaged over the final $10,000 r_g/c$ just before cooling activation.}
    \label{fig:inflow_eq}
\end{figure}

\begin{figure}[htpb]
    \centering
    \includegraphics[width= \linewidth]{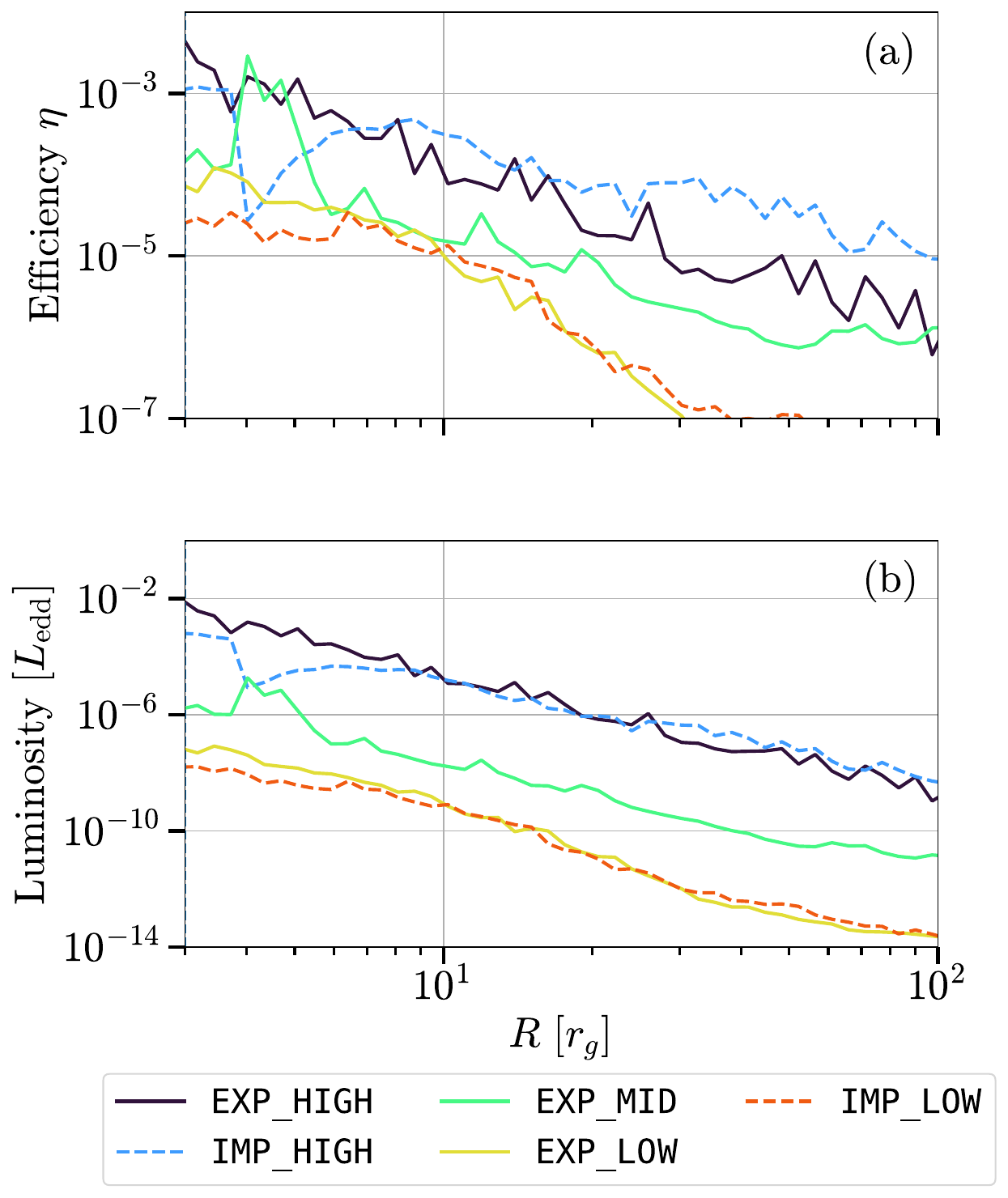}
    \caption{Panel (a): Radial profile of the efficiency as defined by equation \eqref{eq:efficiency_radial} for the last snapshot of each simulation. Panel (b): The luminosity radial profile in Eddington units of the last snapshot of each simulation. We consider only efficiency and luminosity values for $r > 3 r_g$. We show the luminosities and efficiencies at the last snapshot of every simulation (see Tab.~\ref{tab:simulation_list}). }
    \label{fig:efficiency_luminosity}
\end{figure}

In Figure \ref{fig:inflow_eq}(a) we plot the time-averaged accretion rate as a function of equatorial distance. We average it over the interval $48,675 - 53,675 \ r_g/c$ to account for the same period in all simulations. We also show the accretion rate for our initial condition, \code{INIT}, right before activating cooling, averaged throughout $t = 11,300 - 21,300 \ r_g/c$. 

We observe that inflow equilibrium is reached up to approximately $\sim 40 \,\,r_g$ for the \code{INIT} simulation.
We show that all simulations achieve inflow equilibrium up to at least $\sim 80 \ r_g/c$.
In Figure \ref{fig:efficiency_luminosity}(a), we show the radial profile of the radiative efficiency, which is calculated similar to equation \eqref{eq:efficiency_value} following
\begin{equation}
    \eta(r) = \frac{\int_0^{2\pi} \int_0^\pi \sqrt{-g} \ Q^-_{\rm Total} d\theta d\varphi}{\langle \dot{M}c^2 \rangle} \Delta r,
    \label{eq:efficiency_radial}
\end{equation}
where $\Delta r$ is the radial length of the cell. As expected, the efficiency values drop as the accretion rate decreases. However, \code{EXP\_MID} seems to approach \code{EXP\_HIGH} in the very few inner radii $r \approx 4 r_g$, but the overall radial integral to calculate the efficiency values (eq. \ref{eq:efficiency_value}) mitigates the observed spike in \code{EXP\_MID}. Finally, in figure \eqref{fig:efficiency_luminosity}(b), we plot the radial profile of the luminosity calculated as $L = \eta(r)\dot{M}c^2$ in units of Eddington. We observe that \code{EXP\_HIGH} remains beneath the Eddington limit for the radii range considered. Similarly to the efficiency, we also see the expected drop of luminosity as we decrease the accretion rate.

\begin{table*}[htbp]
\centering
\begin{adjustbox}{width=0.65\textwidth}
\begin{tabular}{|c||c|c|c|c|c|}
 \hline
 \multicolumn{6}{|c|}{Simulation List} \\
 \hline
  Name & $\rho_{\text{scale}}$\rule[-2ex]{0pt}{2ex} & \text{Rad. Efficiency} ($\eta$)& $\dot{M} (\dot{M}_{\text{edd}})$ & Time span ($r_g/c$) & Resolution\\
 \hline
 EXP\_HIGH   & $3 \times 10^{-7}$\rule[-2ex]{0pt}{2ex}  &0.15 & $ 2.6 \times 10^{-1}$ & 21300 - 86600 & $384 \times 300 \times 64$ \\
  IMP\_HIGH  & $3 \times 10^{-7}$\rule[-2ex]{0pt}{2ex}& 0.14 & $1.7 \times 10^{-1}$ & 21300 - 68000 & $384 \times 300 \times 64$ \\
 EXP\_MID   & $1 \times 10^{-8}$\rule[-2ex]{0pt}{2ex} & $4.8\times 10^{-2}$ & $5.1 \times 10^{-4}$\rule[-2ex]{0pt}{2ex} & 21300 - 53675 & $384 \times 300 \times 64$ \\
 EXP\_LOW   & $1 \times 10^{-9}$\rule[-2ex]{0pt}{2ex} & $6.6 \times 10^{-3}$ &$ 4.7 \times 10^{-6}$\rule[-2ex]{0pt}{2ex} & 21300 - 53675 & $384 \times 300 \times 64$ \\
 IMP\_LOW   & $1 \times 10^{-9}$\rule[-2ex]{0pt}{2ex} & $4.4 \times 10^{-3}$&$3.0 \times 10^{-6}$ & 21300 - 55325 & $384 \times 300 \times 64$ \\
 \hline
\end{tabular}
\end{adjustbox}
\caption{List of simulations with their parameters. The parameter $\rho_{\rm scale}$ scales the maximum normalized density ($\rho = 1$) to its corresponding value in $\rm g/cm^3$}
\label{tab:simulation_list}
\end{table*}

\subsection{Accretion flow collapse}    \label{sec:flow_thin}

To investigate the effects of cooling on the overall disk structure, we calculate the time-averaged aspect ratio $H/R$ as
\begin{equation}     \label{HoverR}
    \frac{H}{R} = \frac{\int^{2\pi}_0 \int^\pi_0 \sqrt{- g} \rho \lvert \theta - \langle \theta \rangle_\rho \rvert d\theta d\varphi}{\int^{2\pi}_0 \int^\pi_0 \sqrt{- g} \rho d\theta d\varphi},
\end{equation}
where the density-averaged $\langle \theta \rangle_\rho$ is given by
\begin{equation}
    \langle \theta \rangle_\rho = \frac{\int^\pi_0 \sqrt{- g} \rho \theta d\theta}{\int^\pi_0 \sqrt{-g} \rho d\theta}.
    \label{eq:theta_mid}
\end{equation}
We consider $\lvert \theta-\langle \theta \rangle_\rho \lvert$ instead of the standard $\lvert \theta - \pi/2\rvert$ in order to account for oscillations of the disk across the midplane. 

\begin{figure}
    \centering
    \includegraphics[width= \columnwidth]{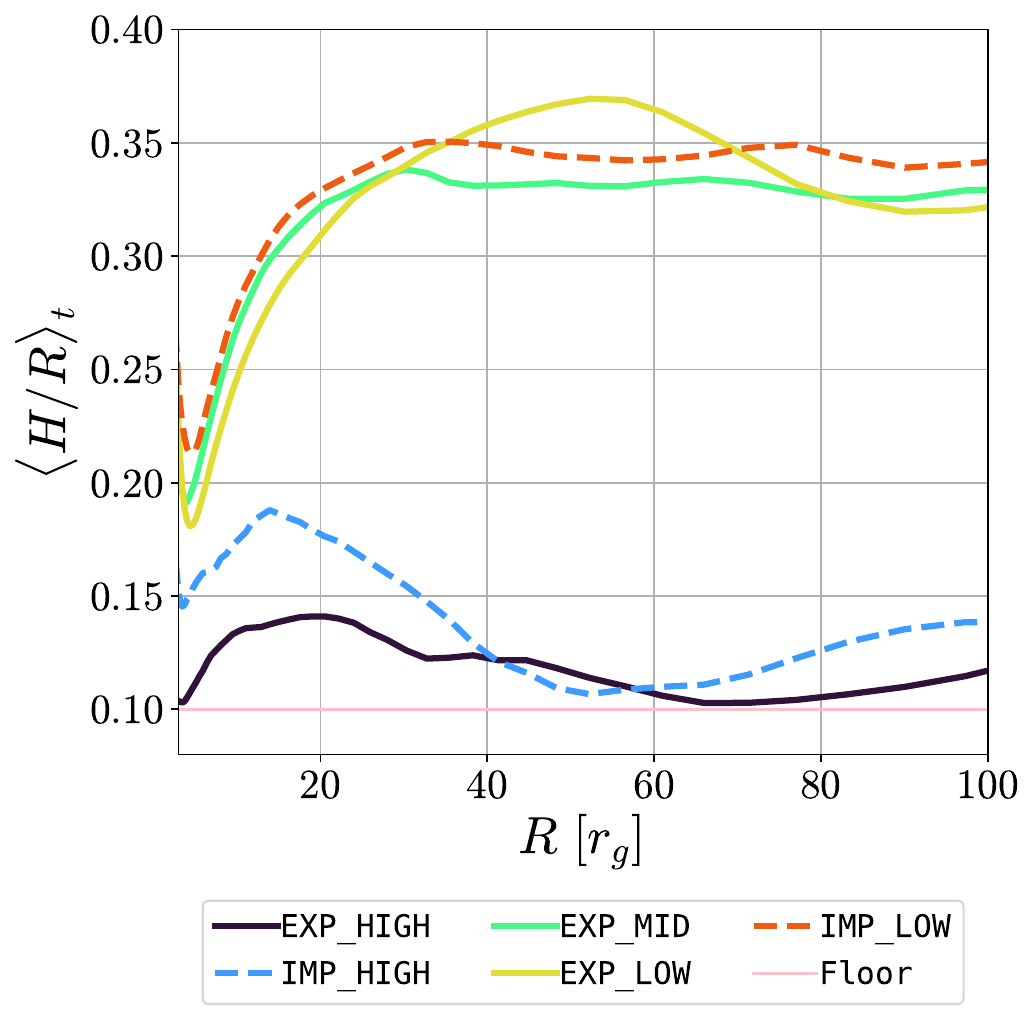}
    \caption{Disk aspect ratio $H/R$ as a function of the distance for all the simulations in Table \ref{tab:simulation_list}, time-averaged from $t = 48,675$ to $53,675 \ r_g/c$.}
    \label{fig:hr_time}
\end{figure}

In Figure \ref{fig:hr_time}, we show $H/R$ averaged over the interval $48,675-53,675 r_g/c$. The disk geometric thickness, $H/R$, defined in Eq.~(\ref{HoverR}), should not be confused with the disk thermal thickness, $H_{\rm th}$, defined in Eq.~(\ref{eq:H_thermal}). Unlike the thermal thickness, the geometric thickness can also be influenced by magnetic forces acting on the accretion flow, particularly in the thin MAD regime \citep{scepi_magnetic_2023,lowell_evidence_2025}.

We choose to average all models over the same time interval for consistency. We see in the figure that for the low $\dot{m}$ models, $H/R$ settles at $\approx 0.35$ for $r \gtrsim 20 r_g$ whereas $H/R \approx 0.2$ for $r \sim r_H$, where $r_H$ is the event horizon radius. This is consistent with other RIAF simulations performed using an adiabatic equation of state \eg{Narayan_2012}. 

As the accretion rate increases, we expect ${H}/{R}$ to decrease due to increased cooling and loss of gas pressure support. When  $\dot{m}$ gets close to a critical value, we see a dramatic shift: the flow vertically collapses, with the lower limit dictated by $h_{\rm floor}$. For these high $\dot{m}$ models, ${H}/{R}$ reaches $\sim 0.16$ in the \code{EXP\_HIGH} run and $\sim 0.23$ for \code{IMP\_HIGH} run. In both models, the disk gets thinner at $r > 50 r_g$. If it were not for our setting of $h_{\rm floor}$, the regions at $r > 50 r_g$ would collapse even further, leading to a much thinner disk. In Section \ref{Comparison between implicit and explicit methods}, we demonstrate the collapse of the accretion disk to our temperature floor for both our high-Eddington simulations (see Fig.~\ref{fig:hr_time}). We conclude that the small contrast in $H/R$ between the thin outer disk and the inner, hot corona-like disk is due to our high $h_{\rm floor}$ which is limited by our choice of numerical resolution. Higher-resolution simulations will provide better insights into the disk thickness contrast in truncated disks.

\begin{figure}
    \centering
    \includegraphics[width= \columnwidth]{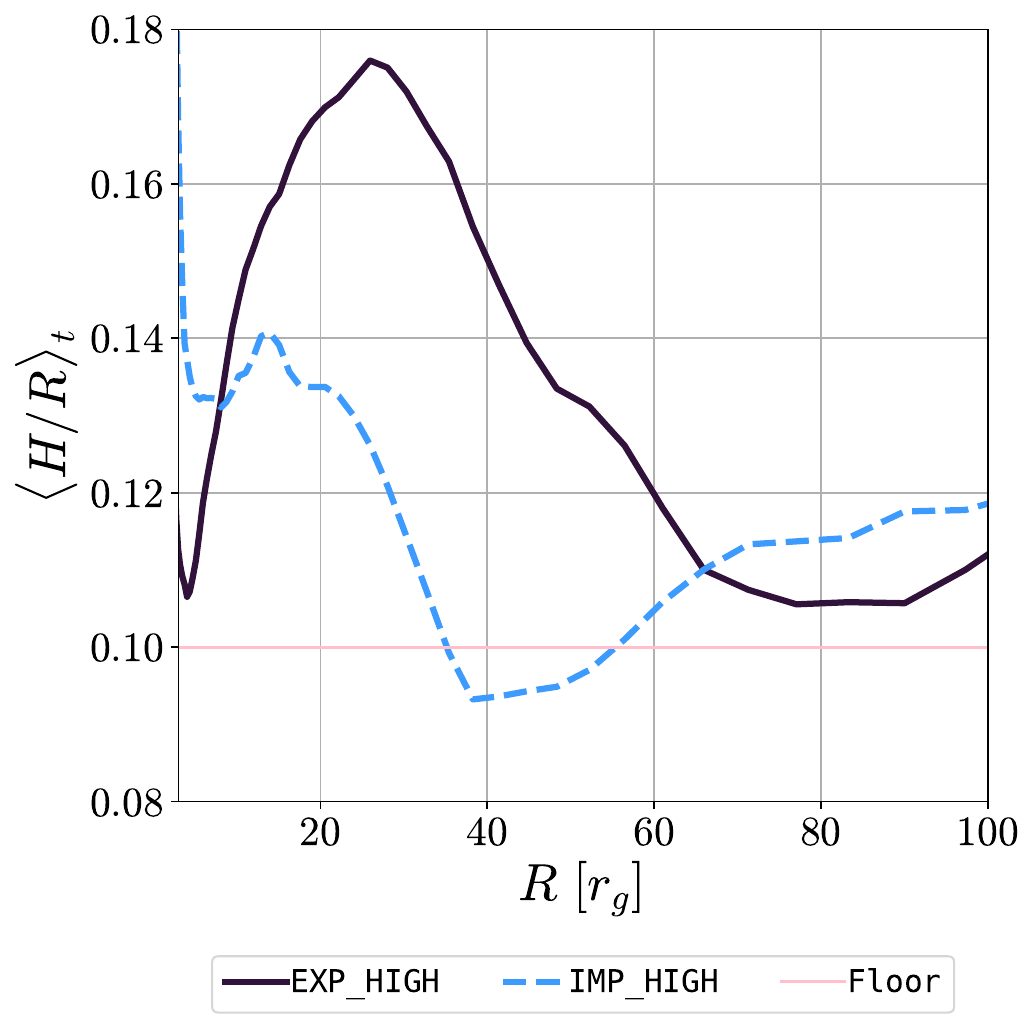}
    \caption{Time-averaged $H/R$ as a function of the equatorial distance for \code{IMP\_HIGH} and \code{EXP\_HIGH} spanning $63,000 -68,000 \ r_g/c$  and $81,600 - 86,600 \ r_g/c$, respectively.}
    \label{fig:hr_highed}
\end{figure}

We investigate the disk thickness further by plotting $H/R$ averaged over the last $5,000 \ r_g/c$ of the higher accretion rates simulations in Figure \ref{fig:hr_highed}. This corresponds to $63,000 -68,000 \ r_g/c$ for \code{IMP\_HIGH} and $81,600 - 86,600 \ r_g/c$ for \code{EXP\_HIGH}, respectively. There is a noticeably higher $H/R$ in the inner regions---$r \lesssim 40r_g$ for the implicit simulation and $r \lesssim 60 r_g$ for the explicit one---although \code{EXP\_HIGH} reaches relatively close to the floor value near the event horizon. The larger thickness of \code{EXP\_HIGH} compared to \code{IMP\_HIGH} is consistent with the radiative efficiency measured at that time (see Fig.~\ref{fig:efficiency_luminosity}). Specifically, \code{IMP\_HIGH} exhibits higher efficiency for $r \gtrsim 10\ r_g$, cooling its structure and resulting in a thinner accretion disk. We note, as discussed in Section \ref{Comparison between implicit and explicit methods}, that the exact value of $H/R$ in the inner regions of the accretion disk ($r < 50 r_g$) depends on the time averaging window, due to the significant time variability of $H/R$.

\begin{figure*}
\centering
\includegraphics[width=\linewidth]{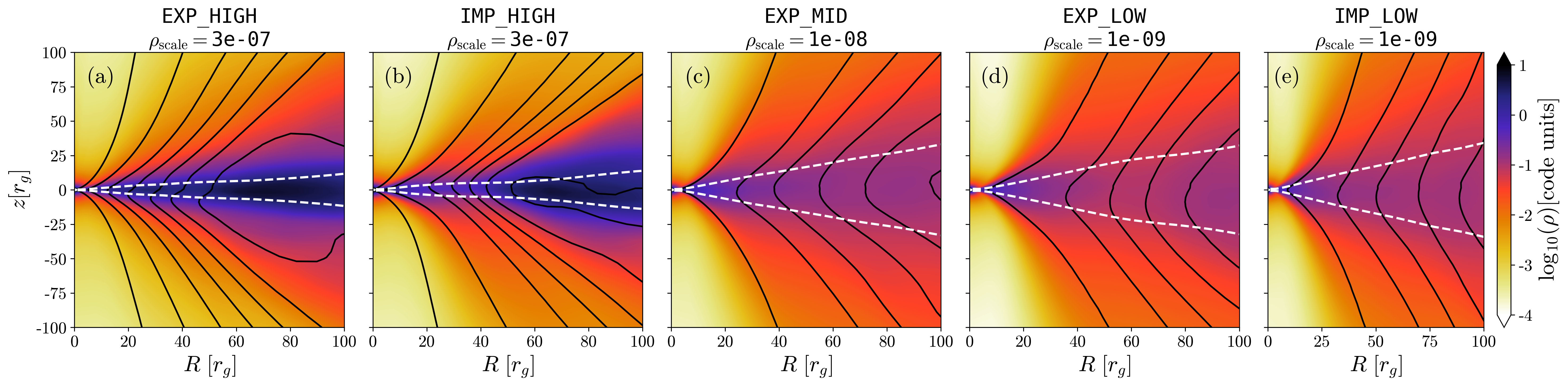}
\caption{Flows at $\dot{m} \sim 0.1$ naturally develop disk truncations (panels a,b), whereas flows at $\dot{m} \ll 0.01$ do not (panels, c-e).  The truncation can be distinguished by the constant $H(r)$, meaning that $H/R$ decreases with radius, signaling the transition to the thin disk. The logarithm of the density, axisymmetrized and time-averaged over the interval $48,675 - 53,675 \ r_g/c$, illustrates the collapse and truncation of the accretion flow at higher accretion rates. The solid black lines correspond to the axisymmetric magnetic potential ($A_\varphi$), and the white dashed lines represent the disk aspect ratio $z = H(x)$.}
\label{fig:avg_rho}
\end{figure*}

The type of aspect ratio profile shown in Figures \ref{fig:hr_time} and \ref{fig:hr_highed} for high $\dot{m}$ runs resembles the ``truncated thin disk + RIAF'' structure invoked to explain certain states of XRBs and also low-luminosity AGNs \eg{Chen1989b,barnier_jet_2023,Nemmen2024}. For a clearer picture, we plot $\log \rho$ in Figure \ref{fig:avg_rho} averaged over $\varphi$ and---most importantly---over time in order to average-out the effect of turbulent fluctuations. We can see clearly in this figure that as $\dot{m}$ increases from right to left, a pronounced amount of gas accumulates in the equatorial region at $r \gtrsim 50 r_g$ corresponding to the darker equatorial regions panels (a) and (b) of Fig. \ref{fig:avg_rho}. As the accretion rate and density decrease, the overall thickness of the flow increase from left to right. Furthermore, while $H/R$ dramatically decreases for $r<20 r_g$ in the simulations \code{EXP\_MID}, \code{EXP\_LOW} and \code{IMP\_LOW}, for the same region the gas resists collapse in the remaining high-$\dot{m}$ models. This trend is also seen in Figure \ref{fig:hr_time}. 

We showed that $\phi_{\rm BH}$ varies with the Eddington ratio: low Eddington ratio models exhibit higher dimensionless magnetic flux ($\phi_{\rm BH} \simeq 50$), while high Eddington ratio models show lower dimensionless magnetic flux ($\phi_{\rm BH} \simeq 20$). This behavior is tied to disk thickness, which strongly influences the large-scale magnetic flux. Thinner disks ($H/R \leq 0.1$) correspond to $\phi_{\rm BH} \simeq 20$, while thicker disks exhibit the fiducial value of $\phi_{\rm BH} \simeq 50$ \citep{Tchekhovskoy_2011,Avara2016,scepi_magnetic_2023,lowell_evidence_2025}. High Eddington models generally feature thinner disks near the BH ($H/R \sim 0.1$), resulting in reduced flux on the BH surface, except for simulation \code{IMP\_HIGH}. Notably, \code{IMP\_HIGH} is twice as thick as \code{EXP\_HIGH}, as shown in Fig.~\ref{fig:hr_time}, which is reflected in its higher $\phi_{\rm BH}$.

Taking together the results depicted in Figures \ref{fig:hr_time}, \ref{fig:hr_highed} and \ref{fig:avg_rho}, we see evidence for the collapse of the accretion flow leading to a thin disk, which gets truncated at $r \sim 50 r_g$.

\subsection{Thermal structure}
\label{sec:thermalstructure}

\begin{figure*}
    \centering
    \includegraphics[width= 0.9\linewidth]{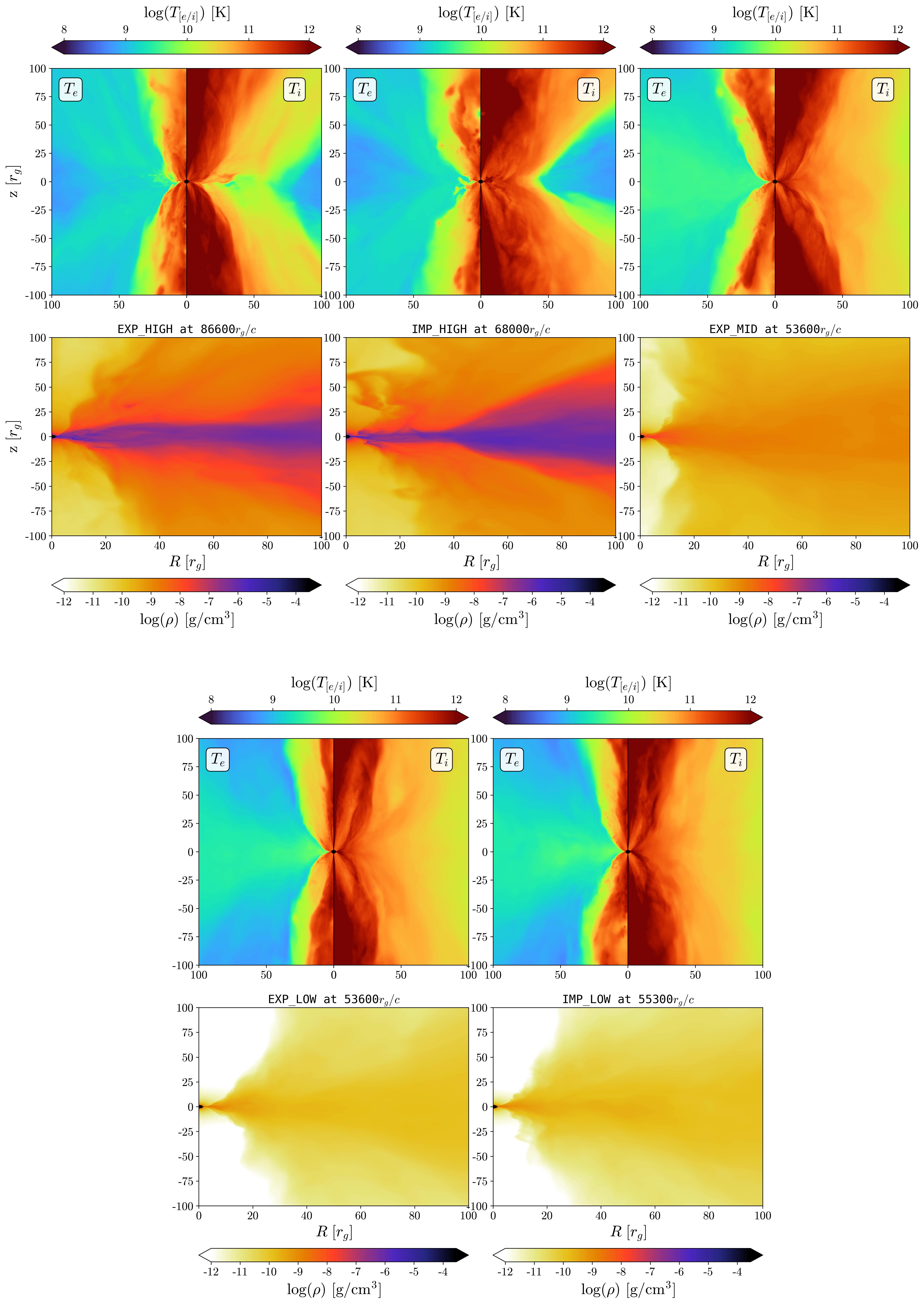}
    \caption{Color maps indicating the instantaneous $\varphi-$averaged electron and ion temperatures in Kelvin, and gas density in $\rm g/cm^3$ for all models at their respective last snapshot.}
    \label{fig:general_grid_plot}
\end{figure*}

Figure \ref{fig:general_grid_plot} shows the slice $\varphi-$averaged of mass density $\rho$, and electron and ion temperatures, $T_{\rm e}$ and $T_{\rm i}$ at the last instant of each simulation (cf. Table \ref{tab:simulation_list}). Looking at the instantaneous density maps of the first two models, \code{EXP\_HIGH} and \code{IMP\_HIGH}, we see a structure resembling a geometrically thin accretion flow. 

Upon inspecting the temperature maps of Figure \ref{fig:general_grid_plot}, we notice a cooler disk with a hot inner corona ($r \lesssim 60r_g$) in the \code{EXP\_HIGH} and \code{IMP\_HIGH} models. Additionally, we notice that, for the same models, the electron and ion temperatures tend to approach each other as the disk becomes geometrically thinner. This occurs at large radii ($r \sim 60r_g$) as seen in Figure \ref{fig:general_grid_plot}. While models with lower Eddington ratios have $T_e<<T_i$ for all radii.
{We also notice a difference in the inner regions ($r \lesssim 60r_g$) of the $\varphi-$averaged temperatures colormap for \code{EXP\_HIGH} and \code{IMP\_HIGH}. \code{EXP\_HIGH} simulation appears to have a much colder ion temperature compared to \code{IMP\_HIGH}, which could be related to the higher $\phi_{\rm BH}$ value, seen in Figure \ref{fig:mdot_phimag}.}

\begin{figure*}
    \centering
    \includegraphics[width= \linewidth]{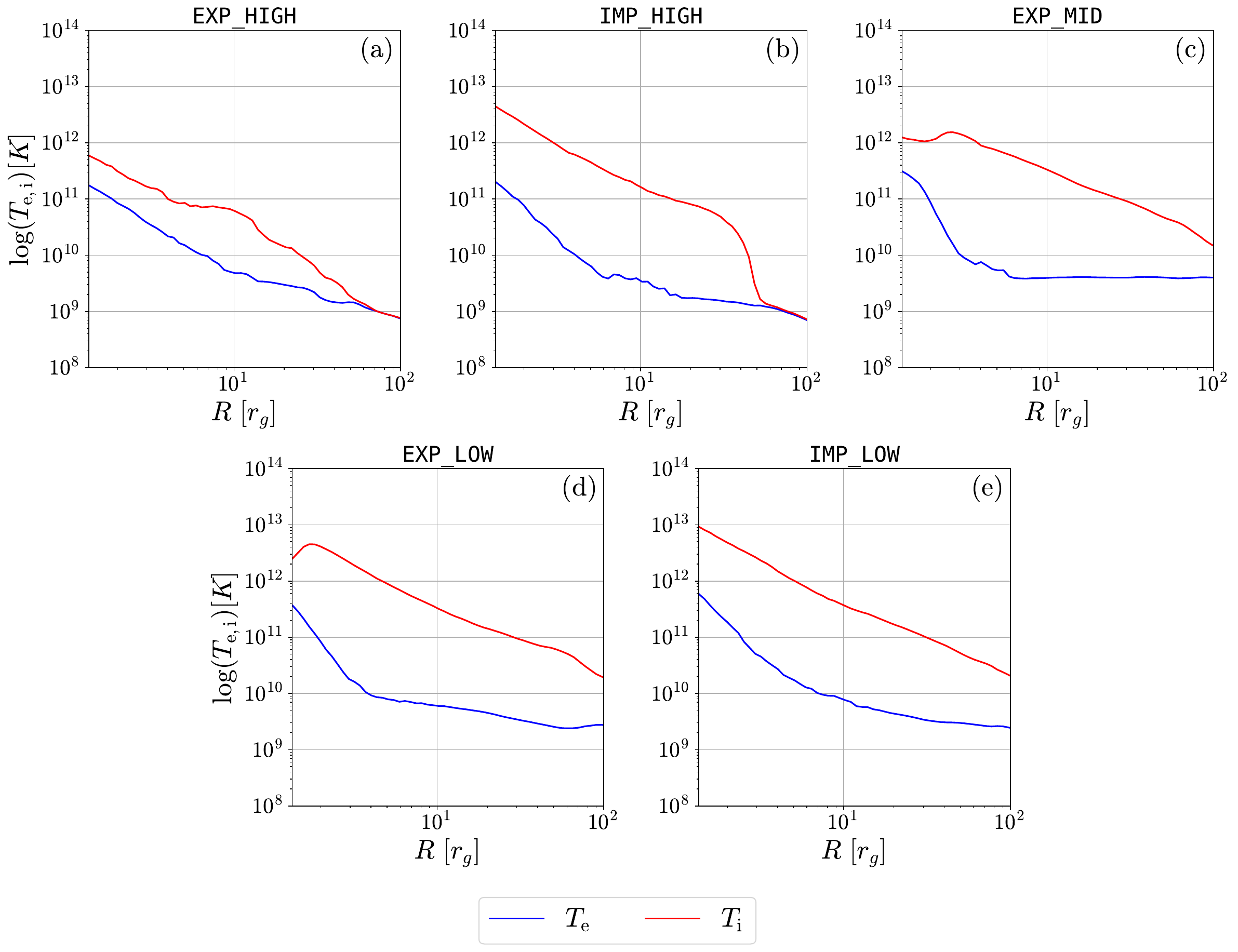}
    \caption{Panels (a)-(e): Electron and ion temperatures averaged over $t$ and $\varphi$ at the $\rho-$averaged poloidal angle $\langle \theta \rangle_\rho$ (eq. \ref{eq:theta_mid}) from $t = 48,675$ to $53,675 r_g/c$ as a function of the distance, for all simulations.}
    \label{fig:temperature_plot}
\end{figure*}

For a clearer view of the difference between the temperatures of the electron and ion populations, we present in Figure \ref{fig:temperature_plot}, the values of $T_{\rm e}$ and $T_{\rm i}$ averaged over $t$ and $\varphi$ and measured at $\theta = \langle \theta\rangle_\rho$ (eq. \ref{eq:theta_mid}). Low $\dot{m}$ simulations \code{IMP\_LOW}, \code{EXP\_LOW} and \code{EXP\_MID} are typified by a two-temperature plasma with ion temperatures consistent with the virial temperature. The temperature difference between electrons and ions is most pronounced in \code{IMP\_LOW} ($T_{\rm i}/T_{\rm e} \approx 55$). In the high-$\dot{m}$ models \code{EXP\_HIGH} and \code{IMP\_HIGH}, though, the inner regions at $r \lesssim 50 r_g$ consist of a two-temperature plasma, while the outer regions become one-temperature with $T_{\rm i}=T_{\rm e} \sim 10^9$ K.  

The onset of a one-temperature plasma occurs due to increased Coulomb collisions in denser regions, which facilitate more energy exchange between ions and electrons. 

In a truncated disk, the temperature is expected to drop from $\sim 10^9 K (\sim 100\ \rm{keV})$ in the inner, optically thin region to $\sim 10^7 K (\sim 1\ \rm{keV})$ in the outer, optically thick region \citep{Novikov1973, Remillard_2006}. However, as shown in Figure \ref{fig:temperature_plot}, the temperature in the outer region is above this expected range by 2 orders of magnitude. This discrepancy arises because our simulations do not take radiation pressure into account. Radiation pressure should become more prominent in the optically thick regions of the disk, where dense matter often interacts with radiation.

\begin{figure}
    \centering
    \includegraphics[width=\linewidth]{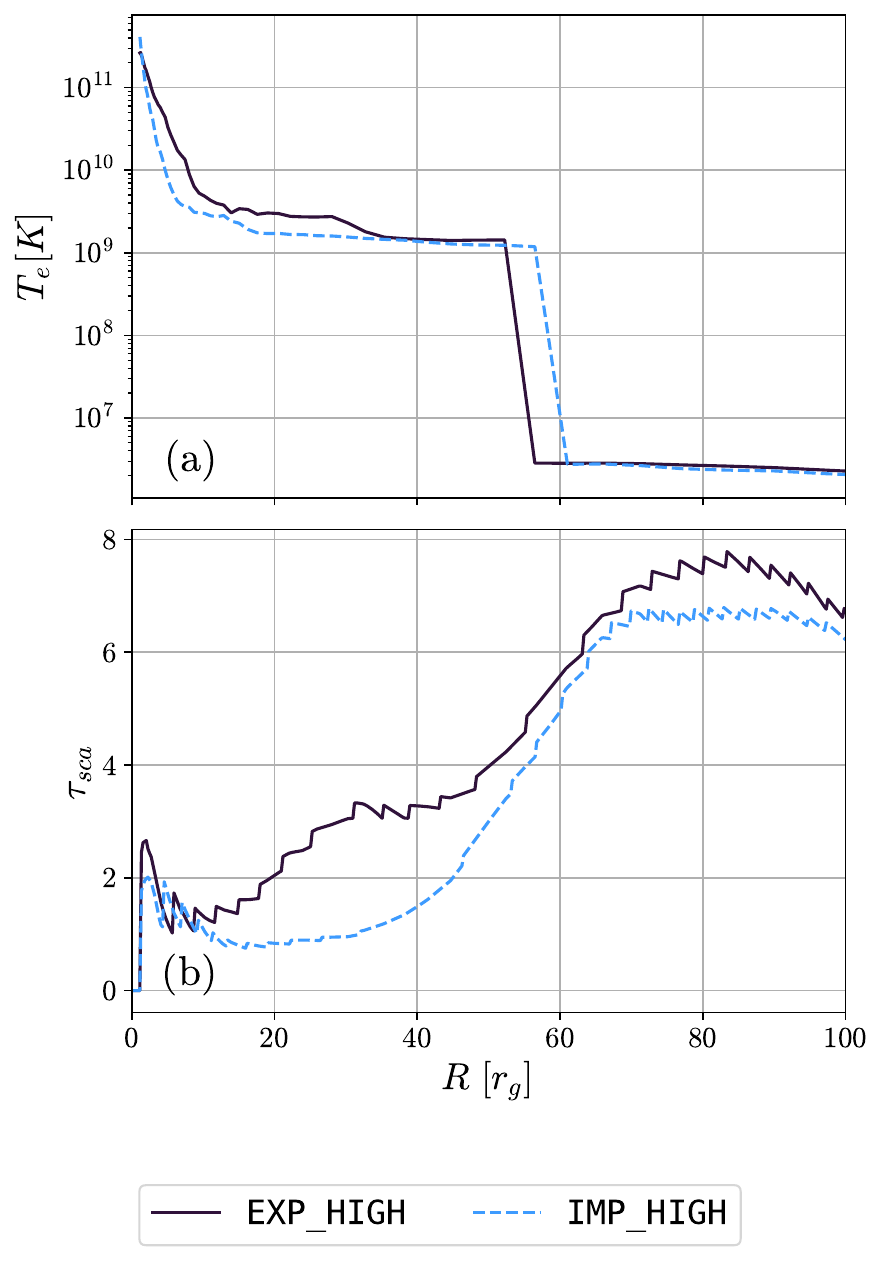}
    \caption{Panel (a): Radial profile of $\tau_{\rm sca} = n_{\rm e} \sigma_{\rm T} H$ for \code{EXP\_HIGH} and \code{IMP\_HIGH}. The dashed line represents the ISCO radius ($r_{\rm ISCO} = 2.04 r_g$ for $a_* = 0.9375$). Panel (b): Radial profile of $T_{\rm e}$ by considering equation \ref{eq:total_pressure} for regions where $\tau_{\rm sca} > 1$. Both images depict the properties averaged from $t = 48,675$ to $53,675/ r_g/c$.}
    \label{fig:tau_temperature_profile}
\end{figure}

To distinguish optically thick and thin regions, we calculate the scattering optical depth by integrating the electron density along the $z-$axis:
\begin{equation}
    \tau_{\rm sca} = \frac{\sigma_{\rm T}}{2}\int_{-H}^H n_{\rm e} dz,
\end{equation}
, where we assume the photon is generated near $z=0$ and travels vertically along its optical path until escaping the disk over a scale height distance $H(r)$. We divide by $\frac{1}{2}$ to average photons escaping from the north and south hemispheres. The probability that a photon escapes without scattering is given by $e^{-\tau_{\rm sca}}$. To define the transition between optically thin and thick regimes, we adopt a threshold based on the escaping fraction of radiation. Specifically, we consider a region to be optically thick when less than $1\%$ of the photons escape unscattered, corresponding to $e^{-\tau_{\rm sca}} = 0.01$. This criterion yields a critical scattering optical depth of $\tau_{\rm sca}\sim 4.6$. In Figure \ref{fig:tau_temperature_profile} (a), we plot the radial profile of $\tau_{\rm sca}$ for \code{EXP\_HIGH} and \code{IMP\_HIGH}. We observe a lower scattering optical depth value $\tau_{\rm sca} \sim 2-3$ in the inner regions, transitioning to $\tau \sim 7-8$. This rise in $\tau_{\rm sca}$ agrees with the observed decrease in the density scale height profile (fig. \ref{fig:hr_time}) and the transition from a two-temperature to a one-temperature plasma (fig. \ref{fig:temperature_plot}).

To correct the temperature in the optically thick regions, we recalculate the electron temperature in post-processing by considering the total pressure as
\begin{equation}
    P = P_{\rm gas} + P_{\rm rad} = aT_{\rm e}^4 + Nk_{\rm B} T_{\rm gas},
    \label{eq:total_pressure}
\end{equation}
where $a=7.5646 \times 10^{-15} \ \rm{erg \ cm^{-3} \ K^{-4}}$ is the radiation constant and $T_{\rm gas} = T_{\rm e} + T_{\rm i}$. We recalculate $T_{\rm e}$ by solving the 4th-degree polynomial equation \eqref{eq:total_pressure} where $\tau_{\rm sca} > 1$ and plot its radial profile in Figure \ref{fig:tau_temperature_profile}(b). We clearly see the temperature drop at $r \sim 50 r_g$ from $T_{\rm e} \sim 10^9K$ to $T_{\rm e} \sim 10^6 K$ due to the transition to a optically thick regime, which is what we expect from observations. 

\begin{figure}
    \centering
    \includegraphics[width=\linewidth]{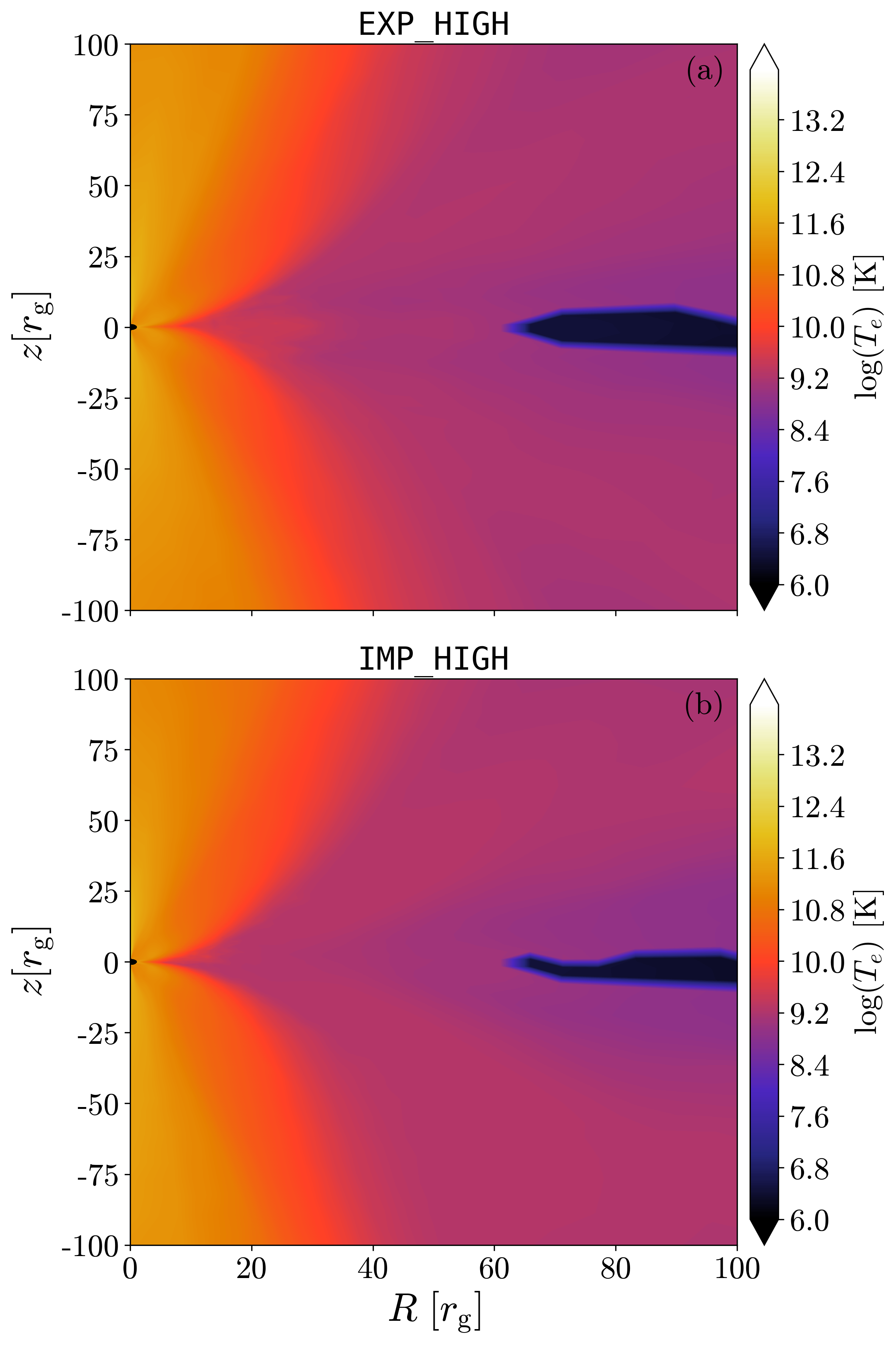}
    \caption{Panels (a)-(b):$(t,\varphi)$-averaged electron temperature colormap for both \code{EXP\_HIGH} and \code{IMP\_HIGH}, respectively. The temperatures, shown in Kelvin, were computed by incorporating radiation pressure in post-processing and solving Equation \eqref{eq:total_pressure}. The gas properties used in this calculation were averaged over the interval $t = 48,675$ to $53,675\ \rm{r_g/c}$.}
    \label{fig:rad_temperature_countour}
\end{figure}

To show the new calculated electron temperature while visualizing the disk structure, we calculate a local in-cell $\tau_{\rm sca}$ value by approximating it as $\tau_{\rm sca}(r,\theta) = n_e(r,\theta) \sigma_{\rm T} H(r)$, where we use similar reasoning as above for using $H$ instead of $2H$. By doing so, we are able to correct the temperature for each cell where the condition $\tau_{\rm sca} > 4.6$ is met. Hence, we plot the $(t,\varphi)$-averaged electron temperature color map in Figure \ref{fig:rad_temperature_countour}. We recover a steeper electron temperature gradient between the inner and the outer regions of the accretion disk. 

In summary, we identify a two-temperature accretion flow spanning the entire flow at low accretion rates, or only the inner puffed-up regions at high accretion rates. We also see a transition from a two to a one-temperature flow at high $\dot{m}$ at $r \approx 50 r_g$, which is also the radii at which the disk thickness changes drastically.

\subsection{Comparison between implicit and explicit methods}
\label{Comparison between implicit and explicit methods}

In Figure \ref{fig:temperature_plot}, the electron and ion temperatures in explicit and implicit simulations show a significant temperature discrepancy in high accretion rate scenarios. For the low accretion rate, implicit and explicit depict a smaller difference, which is expected, since for lower accretion rates, the cooling timescale increases, and higher cooling timescales are easier to handle explicitly. It is important to notice here that our results are subject to the $H/R$ floor of $0.1$.

\begin{figure*}
    \centering
    \includegraphics[width= \linewidth]{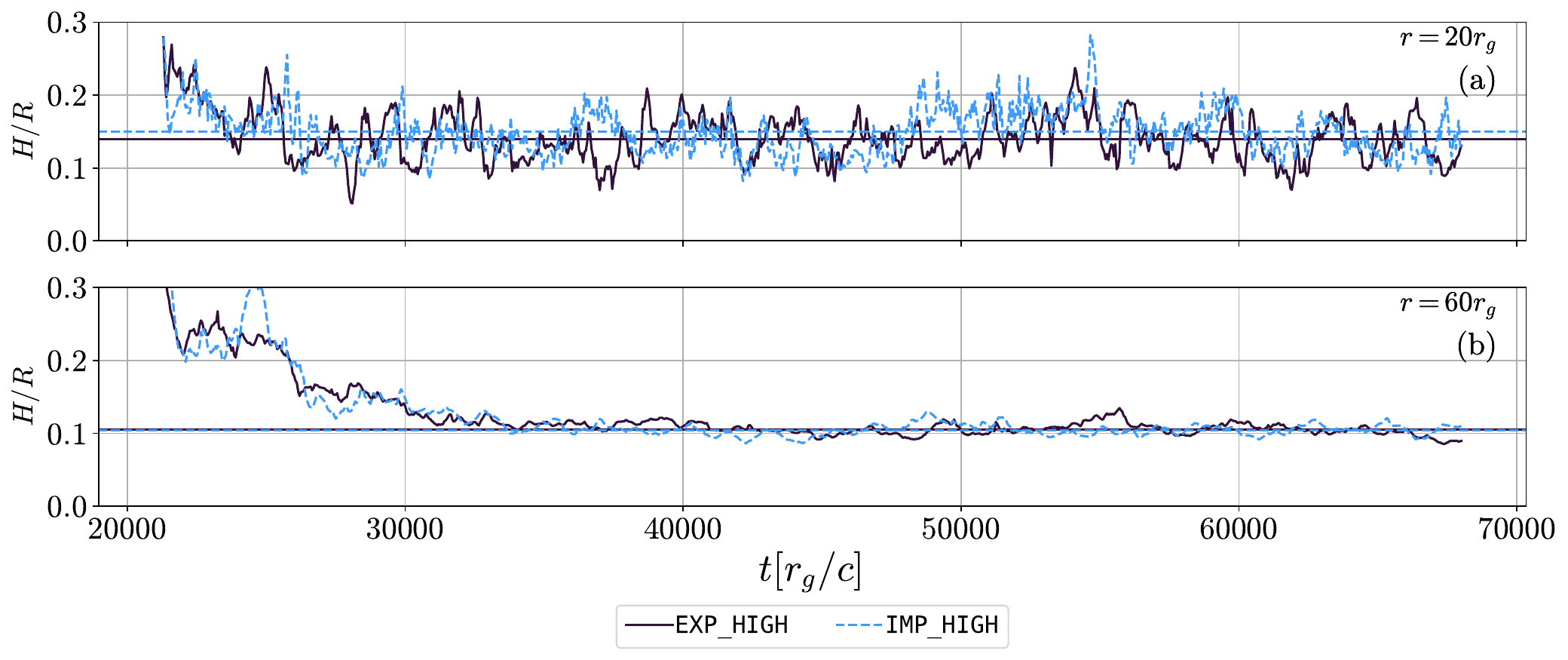}
    \caption{Fig.(a): $H/R$ as a function of time at $r = 20 r_g$ for \code{EXP\_HIGH} and \code{IMP\_HIGH}. The horizontal lines represent the average $H/R$ values for implicit and explicit simulations, with colors corresponding to each simulation type. The averages are computed over the time period $40,000-68,000 r_g/c$. The average values are $H/R \sim 0.14$ for \code{EXP\_HIGH} and $H/R \sim 0.15$ for \code{IMP\_HIGH}. Fig(b): $H/R$ as a function of time at $r = 60 r_g$ for \code{EXP\_HIGH} and \code{IMP\_HIGH}. The dashed lines represent the average $H/R$ values over the period $48,000 - 68,000 r_g/c$ and depict a value near $0.10$ for both simulations.}
    \label{fig:hr_impexp}
\end{figure*}

We compare the evolution of $H/R$ as a function of time for \code{EXP\_HIGH} and \code{IMP\_HIGH} for a radius within the inner, uncollapsed, region ($r = 20 r_g$) and at the outer, collapsed, region ($r = 60 r_g$) in Figure \ref{fig:hr_impexp}(a,b). We first observe that $H/R$ is highly variable in the inner regions, as shown in Fig.\ref{fig:hr_impexp}(b) for $r = 20\,r_g$. These strong oscillations in scale height were noted earlier, where different averaging windows can result in variations of up to a factor of a few in the average $H/R$. Consequently, the difference in $H/R$ between Fig.\ref{fig:hr_time} and Fig.~\ref{fig:hr_highed}, especially for simulation \code{IMP\_HIGH}, can be attributed to this variability.

The simulation \code{IMP\_HIGH} settles at a slightly higher $H/R$ value of $\sim 0.15$ compared to \code{EXP\_HIGH}, which averages $\sim 0.14$ over the period $40k-68k,r_g/c$ in each simulation. These average values are represented by their respective colored horizontal lines. As noted earlier, using a different averaging window could result in a larger discrepancy in $H/R$ between \code{IMP\_HIGH} and \code{EXP\_HIGH} (see Fig.\ref{fig:hr_highed}). 

In Figure \ref{fig:hr_impexp}(b), at an outer radius, both simulations closely align for most of the time. Their time-averaged $H/R$ values, calculated over the period $40k-68k,r_g/c$, converge to approximately $0.10$ for both runs. This convergence occurs because both simulations cool down to the temperature floor defined by $h_{\rm floor}=0.1$. Notably, both simulations approach this limit at a similar rate, suggesting that explicit and implicit cooling methods have minimal impact on the cooling timescale.

Overall, implicit and explicit methods produce very similar results, achieving comparable disk thickness in the inner regions of the accretion flow and collapsing to a thin outer disk at the same rate. However, the implicit method may result in greater variability in disk thickness within the inner regions. The implicit method is approximately $1.8 \times$ slower than the explicit method, though this difference may depend on the resolution and initial conditions (see Appendix \ref{Apendix_sec: efficiency} for further discussion).

\section{Discussion}\label{sec:discussion}

\begin{figure}
    \centering
    \includegraphics[width= \linewidth]{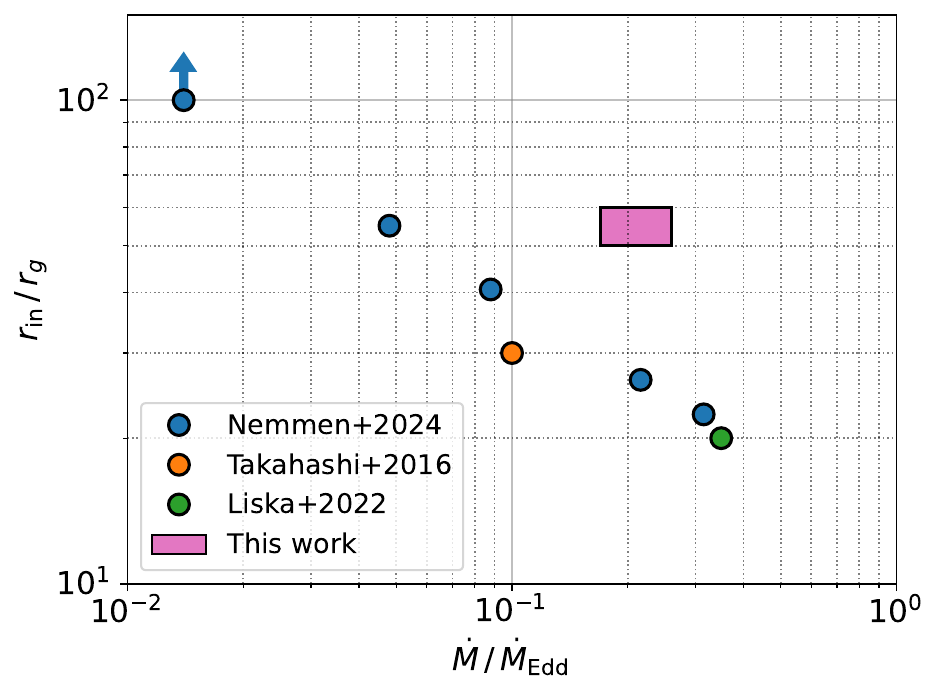}
    \caption{Comparison of accretion disk inner radii as a function of mass accretion rate among different numerical simulations. The square encompasses the range of $\dot{m}$ and $r_{\rm{in}}$ in models \code{IMP\_HIGH} and \code{EXP\_HIGH}.}
    \label{fig:rin}
\end{figure}

We compare the inner radius resulting from our models with other MHD and GRMHD simulations in Figure \ref{fig:rin}. Our values of $r_{\rm{in}}$ are larger than those obtained in other works for the same accretion rates, keeping in mind that our inner radii are rough estimates. It remains to be seen what will be the impact of increased resolution and different magnetic topologies such as SANE. 

Some simulations find no truncation at all, such as those of \citet{Liska_2024} who investigated the effects of cooling in SANE and MAD magnetic topologies using GRRMHD simulations across a broad range of accretion rates. They did not observe disk truncation in either MAD or SANE simulations and suggested that, in the MAD case, where a truncated disk was expected at $r = 40\ r_g$, the simulation probably did not run long enough to reach inflow equilibrium at that radius.

Our temperature values obtained in the midplane generally agree with recent GRMHD simulations of X-ray binaries (XRBs) for the inner hot corona, as reported by \citet{Dihingia_2023}. Observational data suggests that the temperature of the inner optically thick disk in the soft state is around $\sim 10^6 - 10^7 K$ \citep{Islam_2013, Carotenuto_2021}. We found that in \code{EXP\_HIGH} and \code{IMP\_HIGH}, the single-temperature, collapsed outer disk ($r \gtrsim 50\ r_g $) reaches temperatures close to $10^9\ \rm K$, as shown in figure \ref{fig:temperature_plot}. This discrepancy arises from two main factors: (i) the floor value for the disk aspect ratio in our simulations is higher than expected for X-ray binaries in this state, resulting in elevated temperatures, and (ii) the absence of radiation pressure, which depends more steeply on temperature and can support the disk at lower temperatures. We addressed point (ii) in Section~\ref{sec:thermalstructure} by recalculating the temperature using a radiative equation of state in post-processing, assuming radiation pressure was included. This results in significantly lower post-processed temperatures, around  $10^7\ \rm K$.

The ion temperature differs by approximately one order of magnitude between \code{EXP\_HIGH} and \code{IMP\_HIGH}. At $r = 20\ r_g$, the implicit scheme reaches $T_{\rm i} \sim 10^{11} K$, while the explicit scheme results in $T_{\rm i} \sim 10^{10}\ K$, as shown in Figure \ref{fig:temperature_plot}. This temperature difference arises due to vastly different densities and thinner disks as shown in Figure \ref{fig:hr_time}. A denser disk strengthens Coulomb collisions (eq. \ref{eq:coulomb_collisions}), driving the electron and ion temperatures closer together. $\code{IMP\_LOW}$ and $\code{EXP\_LOW}$ do not present a noticeable difference in the temperature profile, as expected, since cooling values are low.

Future work will employ higher-resolution simulations to better resolve the contrast in disk thickness between the inner and outer regions. The focus will be on the transition zone between the thin, one-temperature disk and the thick, two-temperature flow. In this region, large changes in disk thickness are expected to significantly increase the accretion velocity, which depends on disk thickness in the standard framework \citep{shakura_black_1973}. This, in turn, affects the advection velocity of the magnetic field, also linked to disk thickness \citep{jacquemin-ide_magnetic_2021}. Understanding this transition is crucial for unraveling how accretion disks transport mass and magnetic fields and may clarify the conditions driving state transitions in XRBs.

\section{Conclusions} \label{sec:conclusion}

We have investigated a new cooling prescription for GRMHD simulations of accreting black holes, taking into account bremsstrahlung, synchrotron and inverse Compton radiation as well as Coulomb collisions while also leveraging GPU capabilities such as texture memory for fast computations. As a proof of concept, we implemented the cooling procedure, using both explicit and implicit methods, in the \code{H-AMR} code and chose parameters appropriate for a stellar-mass BHs with $a_*=0.9375$ accreting in the MAD low/hard state, starting from the standard torus initial condition. We presented tests of our cooling method and preliminary results of how cooling impacts the accretion flow evolution, varying the mass accretion in the $\sim 10^{-6}-0.1 \dot{M}_{\rm Edd}$ range. Our main results are the following. 

(i) By precomputing the cooling rates and using the GPU texture memory, our approach is $3-5 \times$ faster than radiation M1 closure methods \eg{Sadowski_2013, Sadowski_2014, mckinney2014, Liska_2022} and $2,000$ times faster than numerically solving the relevant equations in a separate test code developed for comparison. 

(ii) Our low $\dot{m}$ models settle at a geometrically thick, low-density, two-temperature hot accretion flow with near Virial temperatures.

(iii) For $\dot{m}$ above a critical rate roughly estimated as $\dot{m}_{\rm crit} \sim 0.01 \dot{m}_{\rm Edd}$, the flow becomes a two-phase medium resembling a colder thin disk and a hot RIAF.

(iv) The inner radius of the thin disk appears to be truncated at $r_{\rm in} \approx 50 r_g$.

(v) In the midplane, the plasma has one temperature at $r>r_{\rm in}$ and two temperatures at $r<r_{\rm in}$.

(vi) Implicit and explicit cooling methods yield similar results, with matching cooling timescales for outer disk collapse and comparable thickness in the inner two-temperature disk.

(vii) Although radiation pressure is not included in the simulation, we analyze its effects in post-processing for the high $\dot{m}$ models. By incorporating it into the equation of state, we correct the temperature in the outer, colder regions of the disk ($r \gtrsim 50\ r_g$), reducing it from $\sim 10^9\ \rm K$ to $\sim 10^7\ \rm K$, in agreement with the expected observational range \citep{Islam_2013, Carotenuto_2021}.

There are three important caveats in our work worth mentioning. The first is that we impose a floor $\min(H/R) = 0.1$ to the accretion thickness to resolve the MRI wavemodes. This floor limits the thin disk resolution attainable here. The second is that our simulations did not evolve for a long enough time to reach inflow equilibrium; to achieve it, we would need to run the models for $\Delta t \sim t_{\rm visc} \sim 10^7 r_g/c$ for the parameters considered in this work. Finally, our set of simulations is by no means exhaustive. Much remains to be explored at higher resolutions, different magnetic topologies (i.e. SANE) and BH spins. 

In the future, our fast GPU cooling calculator can be ported to other GRMHD codes besides \code{H-AMR} and other contexts such as computing neutrino cooling in neutron star mergers.

\begin{acknowledgments}
We acknowledge useful discussions with Nick Kaaz and Beverly Lowell. PNM is grateful to Alexander Tchekhovskoy for his generous hospitality at Northwestern University. PNM acknowledges Ivan Almeida for his assistance with data analysis and simulations. JJ acknowledges Jason Dexter for insightful discussions. 
This work was supported by FAPESP (Fundação de Amparo à Pesquisa do Estado de São Paulo) under grants 2021/01563-5, 2022/07456-9 and 2023/15835-2 (PNM) and 2022/10460-8 (RN).  JJ acknowledges support by the NSF AST-2009884, NASA 80NSSC21K1746 and NASA XMM-Newton  80NSSC22K0799 grants. AT acknowledges support by NASA 
80NSSC22K0031, 
80NSSC22K0799, 
80NSSC18K0565 
and 80NSSC21K1746 
grants, and by the NSF grants 
AST-2009884, 
AST-2107839, 
AST-1815304, 
AST-1911080, 
AST-2206471, 
AST-2407475, 
OAC-2031997. 
This work was performed in part at the Kavli Institute for Theoretical Physics (KITP) supported by grant NSF PHY-2309135.
 An award of computer time was provided by the OLCF Director’s Discretionary Allocation programs under award PHY129.  
Research developed with the help of HPC resources provided by Laboratório Nacional de Computação Científica (LNCC) where we used Sdumont. RN acknowledges NASA support through the \fermi\ Guest Investigator Program (Cycle 16). This work has also used GPUs generously donated by NVIDIA under the GPU Grant Program.
\end{acknowledgments}

\vspace{5mm}

\appendix

\section{Efficiency tests} \label{Apendix_sec: efficiency}

We performed a range of tests to assess the computational efficiency of this prescription in 3D GRMHD simulations. 

\subsection{Time comparison between texture memory and analytical values}

In our study, we undertook a comparative analysis to assess the efficiency of retrieving table values from texture memory as opposed to directly calculating cooling equations. The objective was to quantify the time advantage gained by computing values from a pre-calculated cooling table compared to real-time calculations executed by the code.

\begin{figure}
    \centering
    \includegraphics[width= \columnwidth]{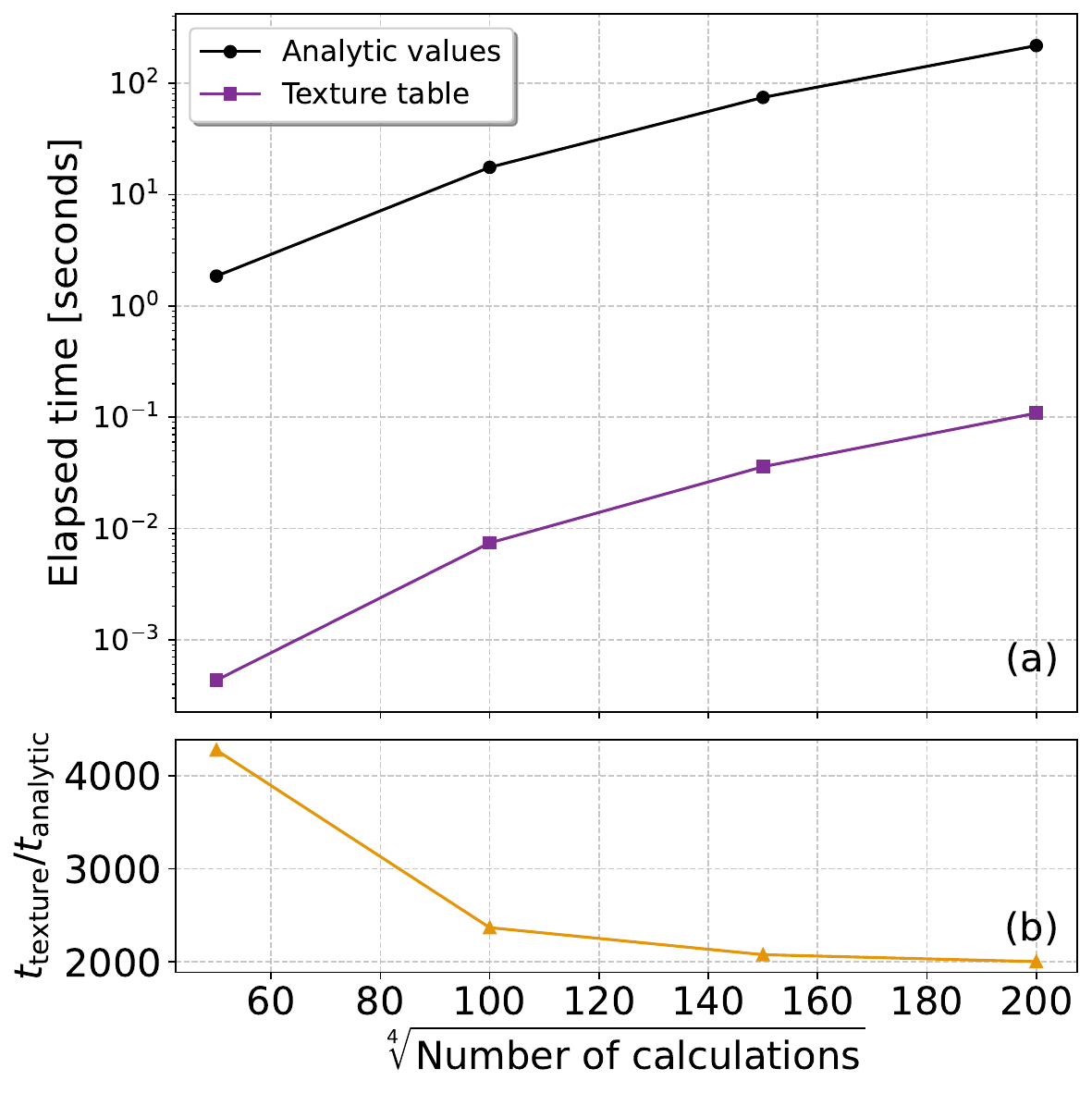}
    \caption{Panel (a): Comparison between texture memory and analytical equations calculation. For this test, we make use of a single GPU NVIDIA QUADRO GP100 using 1792 blocks and 256 threads per block of the GPU. Panel (b): Ratio of elapsed time using texture memory to that of solving the equations analytically.}
    \label{fig:comparison_analytic_texture}
\end{figure}

To conduct this analysis, we isolated the calculations in a separate code from the GRMHD code. We performed computations for $50^4, 100^4, 150^4$, and $200^4$ different values of $Q_{\rm Total}^- (H, B, n_{\rm e}, T_{\rm e})$, and measured the time taken by a single NVIDIA QUADRO GP100 GPU to execute these calculations. We calculate the optimal number of blocks that can fit in the available streaming multiprocessors to get the best result out of the GPU and perform the calculations with 1792 blocks and 256 threads per block. The results, depicting the time taken for both analytical and texture calculations, are presented in Figure \ref{fig:comparison_analytic_texture} for reference.

Notably, we observed that as the number of calculations increases, the time ratio of using direct calculations and texture memory seems to converge, showing that texture memory usage can be up to $2,000$ times faster. It is important to note that while the use of the cooling table can achieve a speedup of approximately $2,000$ times compared to computing the cooling equations on the fly, we do not expect this to translate directly to our GRMHD simulations. This is because the cooling calculations are not the most time-consuming part of the code's overall performance.

\begin{figure}
    \centering    \includegraphics[width=\columnwidth]{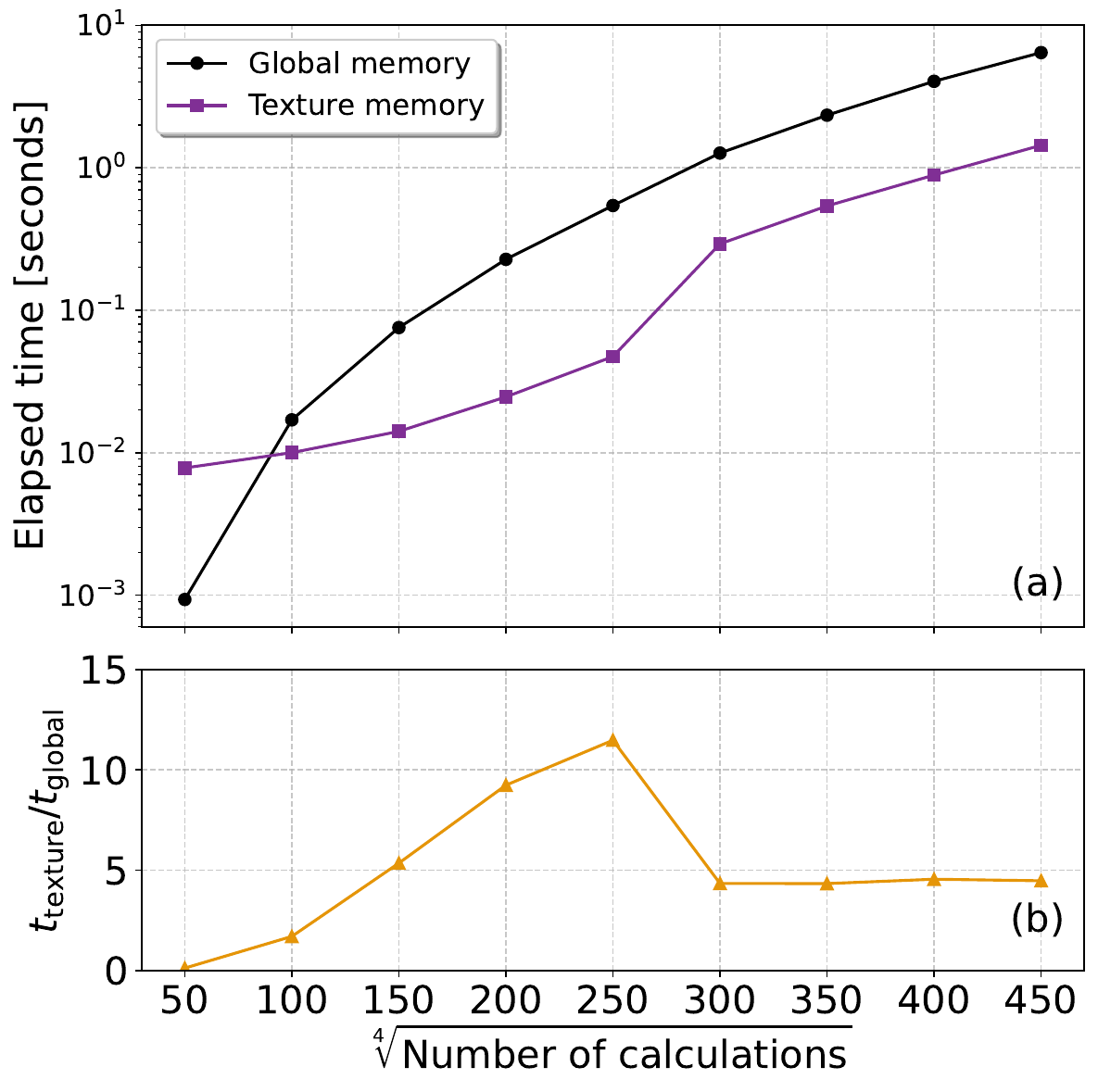}
    \caption{Panel (a): Comparison between texture memory and global memory calculation. For this test, we also make use of a single GPU NVIDIA QUADRO GP100 using 1792 blocks and 256 threads per block of the GPU. Panel (b): The ratio of elapsed time using texture memory to that of using global memory.}
    \label{fig:global_to_texture}
\end{figure}

Additionally, we compare the efficiency of storing these table values in global memory as opposed to texture memory. For this, we performed $50^4$ up to $450^4$ computations and compared the elapsed time between both approaches. We also consider the same GPU and number of threads as in the previous test. We depict the comparison in figure \ref{fig:global_to_texture}.

At first, for a low number of computations, storing the cooling values in global memory seems to be more efficient. However, as the computational workload increases, texture memory provides better performance. It eventually converges to a performance improvement of approximately a factor of $5$ compared to global memory.

\begin{figure}
    \centering    \includegraphics[width=\columnwidth]{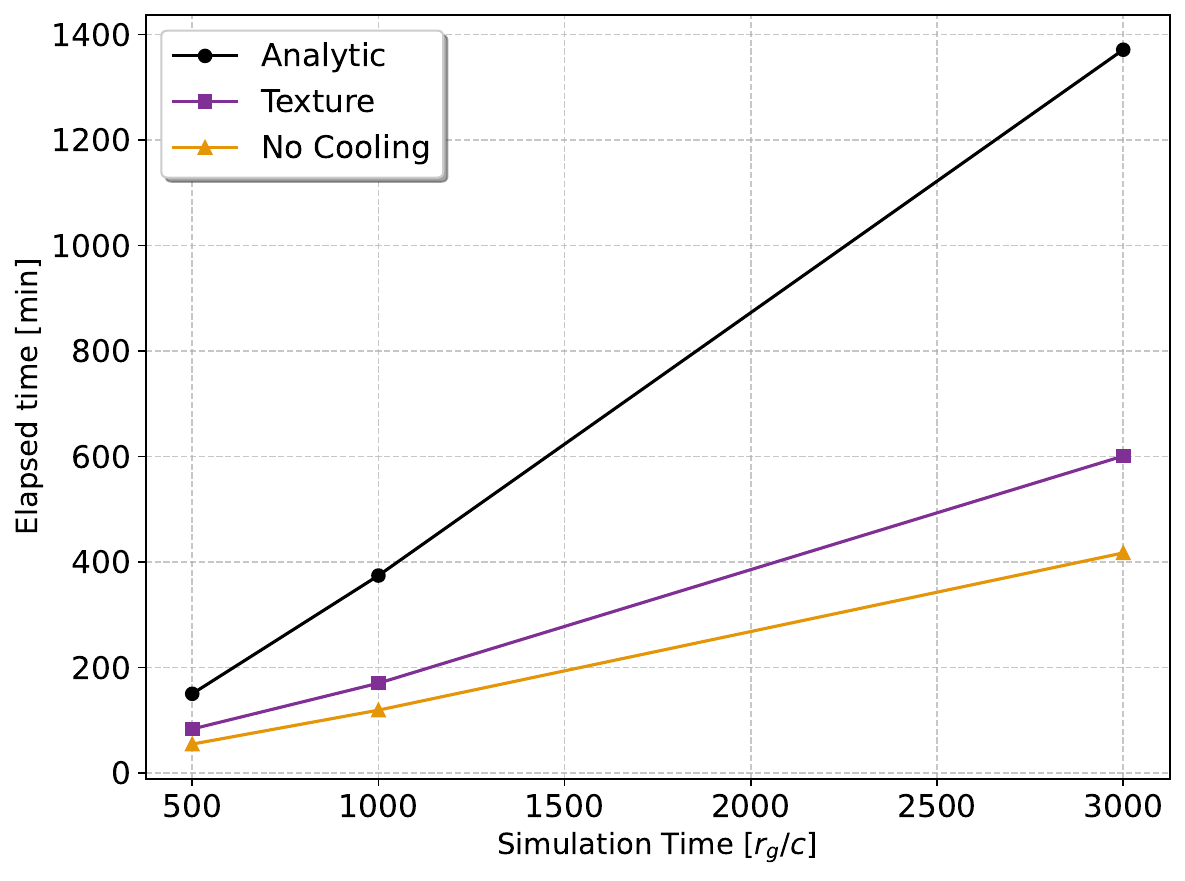}
    \caption{Comparison between texture memory and analytical equations calculation inside \code{H-AMR}. For this test, we make use of a single GPU NVIDIA V100 to simulate with a resolution of $(392 \times 128 \times 112)$. }
    \label{fig:comparison_analytic_texture_inhamr}
\end{figure}

Finally, we conducted a comparative analysis between employing the texture memory look-up table and solving analytical equations in real-time within \code{H-AMR}. To do so, we conducted a 3D simulation with a resolution of $(392 \times 128 \times 112)$ using a single NVIDIA V100 GPU. The results of this comparison are illustrated in Figure \ref{fig:comparison_analytic_texture_inhamr}.

In \code{H-AMR}, the usage of texture memory has demonstrated an efficiency advantage, exhibiting a performance improvement of $2.3 \times$ compared to solving the analytical cooling equations for the first $3,000 r_g/c$ with little computational cost ($1.5 \times$ slower) when compared to the non-cooled system. We expect this efficiency to increase depending on the resolution and duration of the simulation. Since this cooling is calculated within each cell and each timestep, higher resolution systems and longer duration simulations will require more calculations, increasing the difference depicted in figure \ref{fig:comparison_analytic_texture_inhamr}.

\subsection{Comparison to the \code{RAD\_M1} module}

\begin{figure}
    \centering
    \includegraphics[width= \columnwidth]{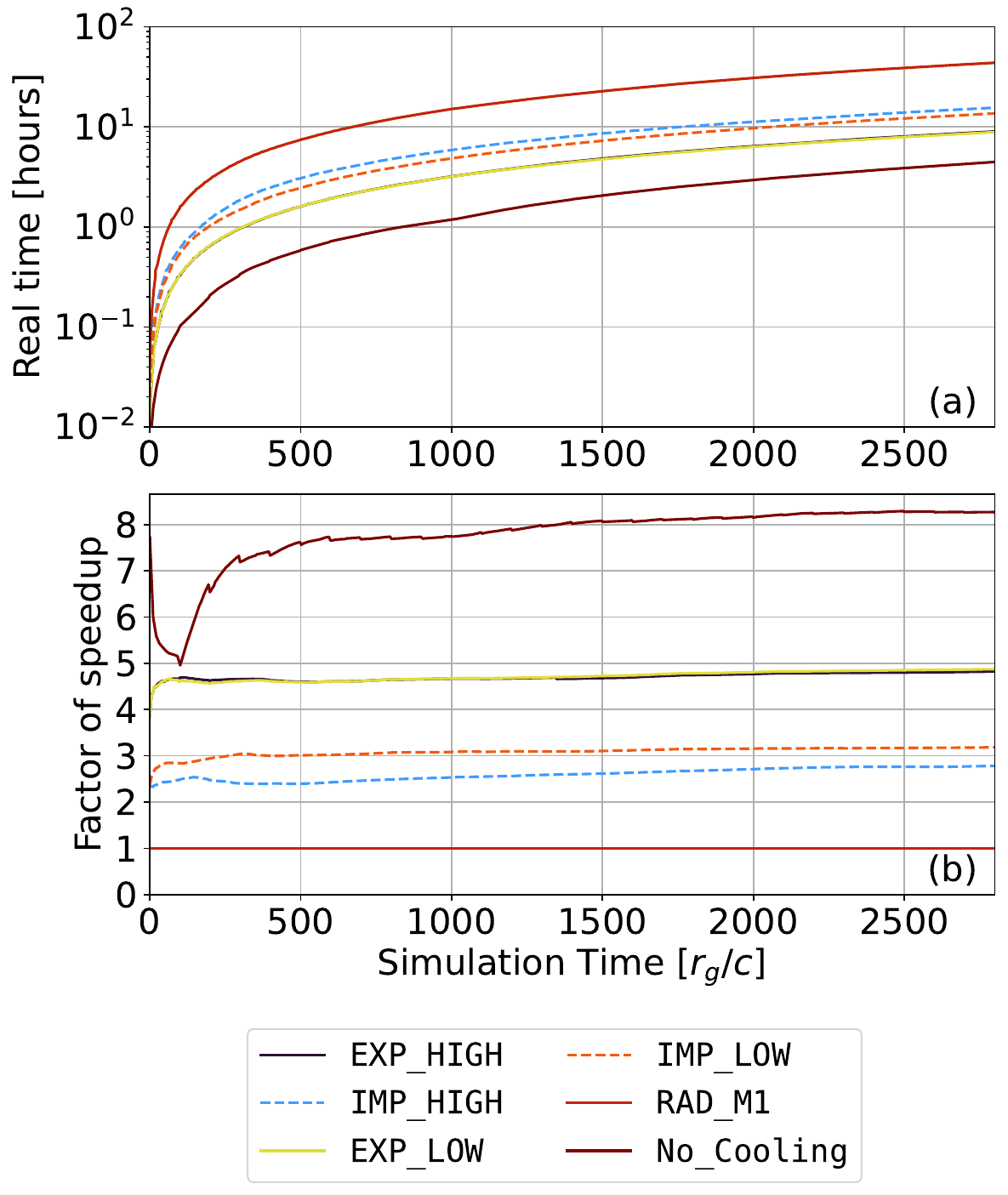}
    \caption{Panel (a): Real-time spent as a function of simulated time, ranging up to 3000 $r_g/c$ after cooling is turned on for all the simulations presented in Table \ref{tab:simulation_list}. Panel (b): Factor of speedup compared to the \code{RAD\_M1} standard module of \code{H-AMR}. We also compare to the initial non-cooled part of the simulations.}
    \label{fig:speedup}
\end{figure}

We compare the time spent to run up to a certain $r_g/c$ for implicit and explicit high and low $\dot{m}$ simulations. We also perform an extra simulation using the standard \code{RAD\_M1} module implemented in \code{H-AMR}. The results are shown in Figure \ref{fig:speedup}. A slowdown of $\approx 1.6\times$ is observed when comparing explicit cooling and no cooling and $\approx 1.8\times$ when comparing explicit and implicit cooling. Additionally, when running the standard \code{RAD\_M1} module of H-AMR, we see a significant slowdown of approximately $8 \times$, $5\times$ and $3\times$ compared to no cooling, explicit cooling and implicit cooling, respectively. We observe a higher impact of the implicit solver for the high accretion rate run, which is expected since cooling is stronger. It is important to note that these values of speedup may vary depending on the accretion rate, resolution, and initial conditions and that we considered the same setup as \code{IMP\_HIGH} for the \code{RAD\_M1}.

\section{Quality Factors}
\label{Appendix:quality_factors}
\begin{figure*}
    \centering    
    \includegraphics[width= \linewidth]{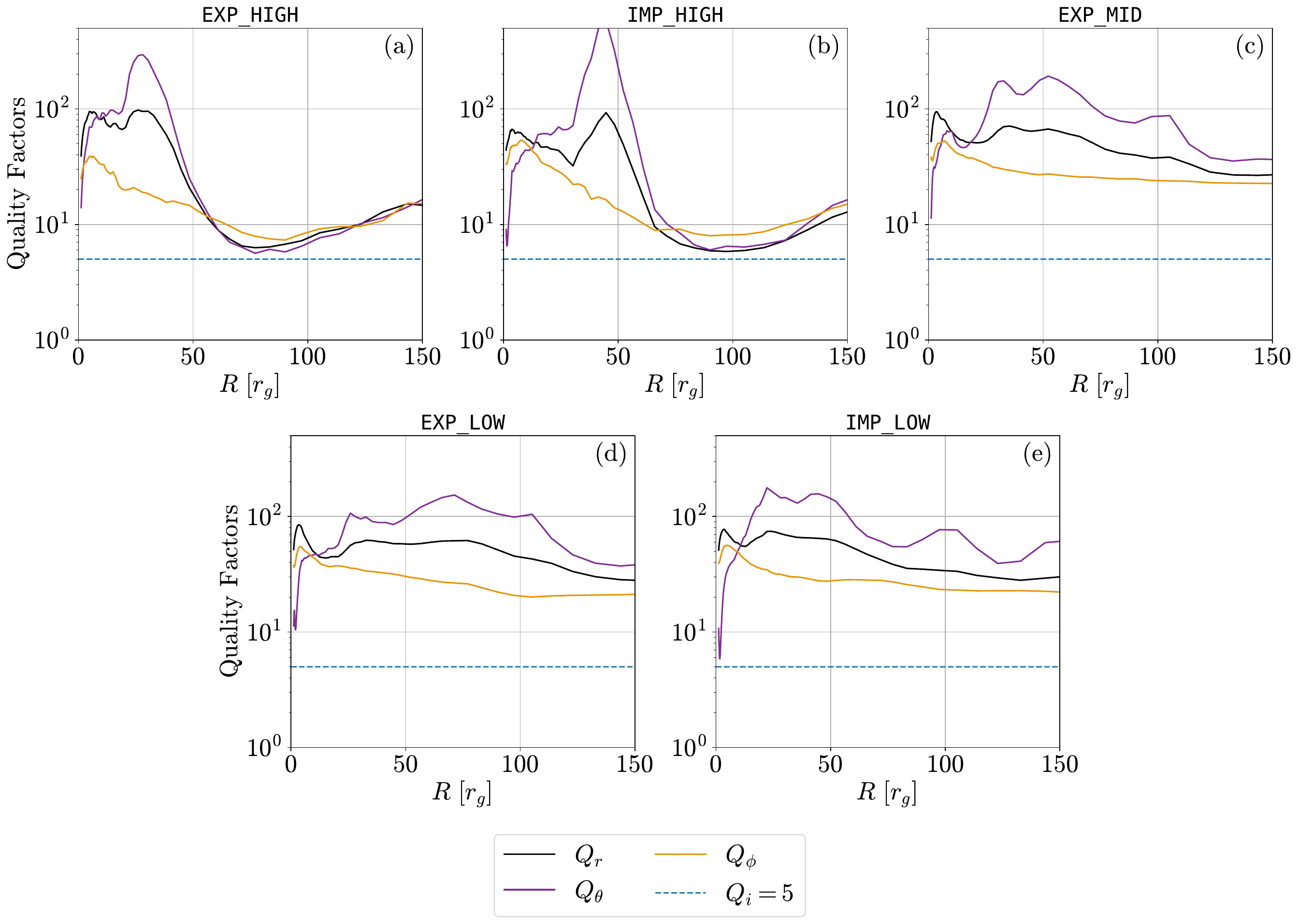}
    \caption{Panels (a)-(e): The radial, poloidal and azimuthal components of the quality factors weighted by $b^2 \rho$ for every simulation described in Table \ref{tab:simulation_list}.}
    \label{fig:qfactors}
\end{figure*}

To guarantee the resolution of the small-scale magnetic processes within the disk, we define a quality parameter based on the characteristic MRI wavelength \citep{Hawley_2011}. 
\begin{equation}
    Q_i = \frac{2 \pi v_A}{\Omega_{rot} dx^i},
\end{equation}
where $v_A = \sqrt{b_\mu b^\mu}/(\rho + B^2 + \gamma \epsilon)$ is the typical alfvén speed, $\Omega_{rot} = \sqrt{u^\mu u^\nu g_{\mu\nu}} \delta^\mu_\nu/u^0$ is the rotational frequency and $dx^i$ is the line element in each direction. We plot $Q_{r}, Q_{\theta}$ and $Q_{\varphi}$ for all the simulations in figure \ref{fig:qfactors}. The established value to capture the evolution of the longest-wavelength MRI modes is within the range of $10-20$. For most regions of the simulation domains, the geometrically thin part of the disk in \code{EXP\_HIGH} and \code{IMP\_HIGH}, approaches a $Q_{r, \theta}$ value of approximately 5.

\clearpage
\bibliography{refs,refsRNS}{}

\end{document}